\pdfoutput=1

\documentclass[12pt,a4paper]{article}

\usepackage{ifthen} 
\newboolean{pdflatex}
\setboolean{pdflatex}{true} 

\newboolean{articletitles}
\setboolean{articletitles}{true} 

\newboolean{uprightparticles}
\setboolean{uprightparticles}{false} 

\newboolean{inbibliography}
\setboolean{inbibliography}{false} 

\usepackage{microtype}
\usepackage{lineno}  
\usepackage{xspace} 
\usepackage{caption} 

\usepackage{graphicx}  
\usepackage{color}
\usepackage{colortbl}
\graphicspath{{./figs/}} 

\usepackage{amsmath} 
\usepackage{amssymb}
\usepackage{amsfonts}
\usepackage{upgreek} 

\newcommand*\patchAmsMathEnvironmentForLineno[1]{%
\expandafter\let\csname old#1\expandafter\endcsname\csname #1\endcsname
\expandafter\let\csname oldend#1\expandafter\endcsname\csname
end#1\endcsname
 \renewenvironment{#1}%
   {\linenomath\csname old#1\endcsname}%
   {\csname oldend#1\endcsname\endlinenomath}%
}
\newcommand*\patchBothAmsMathEnvironmentsForLineno[1]{%
  \patchAmsMathEnvironmentForLineno{#1}%
  \patchAmsMathEnvironmentForLineno{#1*}%
}
\AtBeginDocument{%
\patchBothAmsMathEnvironmentsForLineno{equation}%
\patchBothAmsMathEnvironmentsForLineno{align}%
\patchBothAmsMathEnvironmentsForLineno{flalign}%
\patchBothAmsMathEnvironmentsForLineno{alignat}%
\patchBothAmsMathEnvironmentsForLineno{gather}%
\patchBothAmsMathEnvironmentsForLineno{multline}%
\patchBothAmsMathEnvironmentsForLineno{eqnarray}%
}

\usepackage{hyperref}    
\usepackage[all]{hypcap} 

\def\lhcb {\mbox{LHCb}\xspace}

\def\MagUp {\mbox{\em Mag\kern -0.05em Up}\xspace}

\ifthenelse{\boolean{uprightparticles}}%
{

 \def\Ppi         {\ensuremath{\uppi}\xspace}

 \def\Ppsi        {\ensuremath{\uppsi}\xspace}

 \def\PDelta      {\ensuremath{\Delta}\xspace}                 
 \def\PXi      {\ensuremath{\Xi}\xspace}                 
 \def\PLambda      {\ensuremath{\Lambda}\xspace}                 
 \def\PSigma      {\ensuremath{\Sigma}\xspace}                 
 \def\POmega      {\ensuremath{\Omega}\xspace}                 
 \def\PUpsilon      {\ensuremath{\Upsilon}\xspace}                 
 
 \def\PB      {\ensuremath{\mathrm{B}}\xspace}                 
                  
 \def\PD      {\ensuremath{\mathrm{D}}\xspace}

 \def\PJ      {\ensuremath{\mathrm{J}}\xspace}                 
 \def\PK      {\ensuremath{\mathrm{K}}\xspace}

 \def\Pb      {\ensuremath{\mathrm{b}}\xspace}                 
 \def\Pc      {\ensuremath{\mathrm{c}}\xspace}                 
 \def\Pd      {\ensuremath{\mathrm{d}}\xspace}

 \def\Pi      {\ensuremath{\mathrm{i}}\xspace}

 \def\Ps      {\ensuremath{\mathrm{s}}\xspace}                 
                  
 \def\Pu      {\ensuremath{\mathrm{u}}\xspace}

}
{

 \def\Ppi         {\ensuremath{\pi}\xspace}

 \def\Ppsi        {\ensuremath{\psi}\xspace}                 
                  
 \mathchardef\PDelta="7101
 \mathchardef\PXi="7104
 \mathchardef\PLambda="7103
 \mathchardef\PSigma="7106
 \mathchardef\POmega="710A
 \mathchardef\PUpsilon="7107
                  
 \def\PB      {\ensuremath{B}\xspace}                 
                  
 \def\PD      {\ensuremath{D}\xspace}

 \def\PJ      {\ensuremath{J}\xspace}                 
 \def\PK      {\ensuremath{K}\xspace}

 \def\Pb      {\ensuremath{b}\xspace}                 
 \def\Pc      {\ensuremath{c}\xspace}                 
 \def\Pd      {\ensuremath{d}\xspace}

 \def\Pi      {\ensuremath{i}\xspace}

 \def\Ps      {\ensuremath{s}\xspace}                 
                  
 \def\Pu      {\ensuremath{u}\xspace}

}

\makeatletter
\ifcase \@ptsize \relax
  \newcommand{\miniscule}{\@setfontsize\miniscule{4}{5}}
\or
  \newcommand{\miniscule}{\@setfontsize\miniscule{5}{6}}
\or
  \newcommand{\miniscule}{\@setfontsize\miniscule{5}{6}}
\fi
\makeatother

\DeclareRobustCommand{\optbar}[1]{\shortstack{{\miniscule (\rule[.5ex]{1.25em}{.18mm})}
  \\ [-.7ex] $#1$}}

\def\uquark    {{\ensuremath{\Pu}}\xspace}

\def\dquark    {{\ensuremath{\Pd}}\xspace}

\def\squark    {{\ensuremath{\Ps}}\xspace}

\def\cquark    {{\ensuremath{\Pc}}\xspace}

\def\bquark    {{\ensuremath{\Pb}}\xspace}

\def\pion   {{\ensuremath{\Ppi}}\xspace}
\def\piz    {{\ensuremath{\pion^0}}\xspace}

\def\pip    {{\ensuremath{\pion^+}}\xspace}
\def\pim    {{\ensuremath{\pion^-}}\xspace}

\def\kaon    {{\ensuremath{\PK}}\xspace}
  \def\Kbar    {{\kern 0.2em\overline{\kern -0.2em \PK}{}}\xspace}

\def\KorKbar    {\kern 0.18em\optbar{\kern -0.18em K}{}\xspace}

\def\Kp      {{\ensuremath{\kaon^+}}\xspace}
\def\Km      {{\ensuremath{\kaon^-}}\xspace}

\def\KS      {{\ensuremath{\kaon^0_{\rm\scriptscriptstyle S}}}\xspace}

\def\Kstarz  {{\ensuremath{\kaon^{*0}}}\xspace}
\def\Kstarzb {{\ensuremath{\Kbar{}^{*0}}}\xspace}

  \def\Dbar    {{\kern 0.2em\overline{\kern -0.2em \PD}{}}\xspace}
\def\D       {{\ensuremath{\PD}}\xspace}

\def\DorDbar    {\kern 0.18em\optbar{\kern -0.18em D}{}\xspace}
\def\Dz      {{\ensuremath{\D^0}}\xspace}
\def\Dzb     {{\ensuremath{\Dbar{}^0}}\xspace}
\def\Dp      {{\ensuremath{\D^+}}\xspace}

\def\Dstarz  {{\ensuremath{\D^{*0}}}\xspace}
\def\Dstarzb {{\ensuremath{\Dbar{}^{*0}}}\xspace}
\def\Dstarp  {{\ensuremath{\D^{*+}}}\xspace}
\def\Dstarm  {{\ensuremath{\D^{*-}}}\xspace}

\def\Ds      {{\ensuremath{\D^+_\squark}}\xspace}
\def\Dsp     {{\ensuremath{\D^+_\squark}}\xspace}
\def\Dsm     {{\ensuremath{\D^-_\squark}}\xspace}

\def\B       {{\ensuremath{\PB}}\xspace}
\def\Bbar    {{\ensuremath{\kern 0.18em\overline{\kern -0.18em \PB}{}}}\xspace}

\def\BorBbar    {\kern 0.18em\optbar{\kern -0.18em B}{}\xspace}
\def\Bz      {{\ensuremath{\B^0}}\xspace}
\def\Bzb     {{\ensuremath{\Bbar{}^0}}\xspace}
\def\Bu      {{\ensuremath{\B^+}}\xspace}
\def\Bub     {{\ensuremath{\B^-}}\xspace}
\def\Bp      {{\ensuremath{\Bu}}\xspace}
\def\Bm      {{\ensuremath{\Bub}}\xspace}

\def\Bs      {{\ensuremath{\B^0_\squark}}\xspace}
\def\Bsb     {{\ensuremath{\Bbar{}^0_\squark}}\xspace}

\def\jpsi     {{\ensuremath{{\PJ\mskip -3mu/\mskip -2mu\Ppsi\mskip 2mu}}}\xspace}

  \def\Y#1S{\ensuremath{\PUpsilon{(#1S)}}\xspace}

\def\Lz          {{\ensuremath{\PLambda}}\xspace}
\def\Lbar        {{\ensuremath{\kern 0.1em\overline{\kern -0.1em\PLambda}}}\xspace}
\def\LorLbar    {\kern 0.18em\optbar{\kern -0.18em \PLambda}{}\xspace}

\def\Lb      {{\ensuremath{\Lz^0_\bquark}}\xspace}

\def\to                 {\ensuremath{\rightarrow}\xspace}

\def\CP                {{\ensuremath{C\!P}}\xspace}

\def\Vud  {{\ensuremath{V_{\uquark\dquark}}}\xspace}
\def\Vcd  {{\ensuremath{V_{\cquark\dquark}}}\xspace}

\def\Vus  {{\ensuremath{V_{\uquark\squark}}}\xspace}
\def\Vcs  {{\ensuremath{V_{\cquark\squark}}}\xspace}

\def\AT#1     {\ensuremath{A_{\mathrm{T}}^{#1}}\xspace}           

\def\C#1      {\ensuremath{\mathcal{C}_{#1}}\xspace}                       
\def\Cp#1     {\ensuremath{\mathcal{C}_{#1}^{'}}\xspace}                    
\def\Ceff#1   {\ensuremath{\mathcal{C}_{#1}^{\mathrm{(eff)}}}\xspace}        
\def\Cpeff#1  {\ensuremath{\mathcal{C}_{#1}^{'\mathrm{(eff)}}}\xspace}       
\def\Ope#1    {\ensuremath{\mathcal{O}_{#1}}\xspace}                       
\def\Opep#1   {\ensuremath{\mathcal{O}_{#1}^{'}}\xspace}                    

\newcommand{\tev}{\ifthenelse{\boolean{inbibliography}}{\ensuremath{~T\kern -0.05em eV}\xspace}{\ensuremath{\mathrm{\,Te\kern -0.1em V}}}\xspace}
\newcommand{\gev}{\ensuremath{\mathrm{\,Ge\kern -0.1em V}}\xspace}
\newcommand{\mev}{\ensuremath{\mathrm{\,Me\kern -0.1em V}}\xspace}
\newcommand{\kev}{\ensuremath{\mathrm{\,ke\kern -0.1em V}}\xspace}
\newcommand{\ev}{\ensuremath{\mathrm{\,e\kern -0.1em V}}\xspace}
\newcommand{\gevc}{\ensuremath{{\mathrm{\,Ge\kern -0.1em V\!/}c}}\xspace}
\newcommand{\mevc}{\ensuremath{{\mathrm{\,Me\kern -0.1em V\!/}c}}\xspace}
\newcommand{\gevcc}{\ensuremath{{\mathrm{\,Ge\kern -0.1em V\!/}c^2}}\xspace}
\newcommand{\gevgevcccc}{\ensuremath{{\mathrm{\,Ge\kern -0.1em V^2\!/}c^4}}\xspace}
\newcommand{\mevcc}{\ensuremath{{\mathrm{\,Me\kern -0.1em V\!/}c^2}}\xspace}

\def\mum  {\ensuremath{{\,\upmu\rm m}}\xspace}

\def\invfb   {\ensuremath{\mbox{\,fb}^{-1}}\xspace}

\newcommand{\chisq}{\ensuremath{\chi^2}\xspace}

\newcommand{\chisqip}{\ensuremath{\chi^2_{\rm IP}}\xspace}

\def\gsim{{~\raise.15em\hbox{$>$}\kern-.85em
          \lower.35em\hbox{$\sim$}~}\xspace}
\def\lsim{{~\raise.15em\hbox{$<$}\kern-.85em
          \lower.35em\hbox{$\sim$}~}\xspace}

\def\sPlot{\mbox{\em sPlot}\xspace}

\def\ptot       {\mbox{$p$}\xspace}
\def\pt         {\mbox{$p_{\rm T}$}\xspace}

\def\evtgen     {\mbox{\textsc{EvtGen}}\xspace}

\def\geant      {\mbox{\textsc{Geant4}}\xspace}

\def\photos     {\mbox{\textsc{Photos}}\xspace}

\def\pythia     {\mbox{\textsc{Pythia}}\xspace}

\def\tell1  {TELL1\xspace}
\def\ukl1   {UKL1\xspace}

\def\Asy{{\mathcal{A}}}

\def\brf{{\cal{B}}}

\def\nonc{\cancel{c}}
\def\Dveto{\cancel{D}}

\usepackage{cancel}
\usepackage{cite} 
\usepackage{mciteplus}

\usepackage{longtable} 

\begin{document}

\renewcommand{\thefootnote}{\fnsymbol{footnote}}
\setcounter{footnote}{1}


\begin{titlepage}
\pagenumbering{roman}

\vspace*{-1.5cm}
\centerline{\large EUROPEAN ORGANIZATION FOR NUCLEAR RESEARCH (CERN)}
\vspace*{1.5cm}
\hspace*{-0.5cm}
\begin{tabular*}{\linewidth}{lc@{\extracolsep{\fill}}r}
\ifthenelse{\boolean{pdflatex}}
{\vspace*{-2.7cm}\mbox{\!\!\!\includegraphics[width=.14\textwidth]{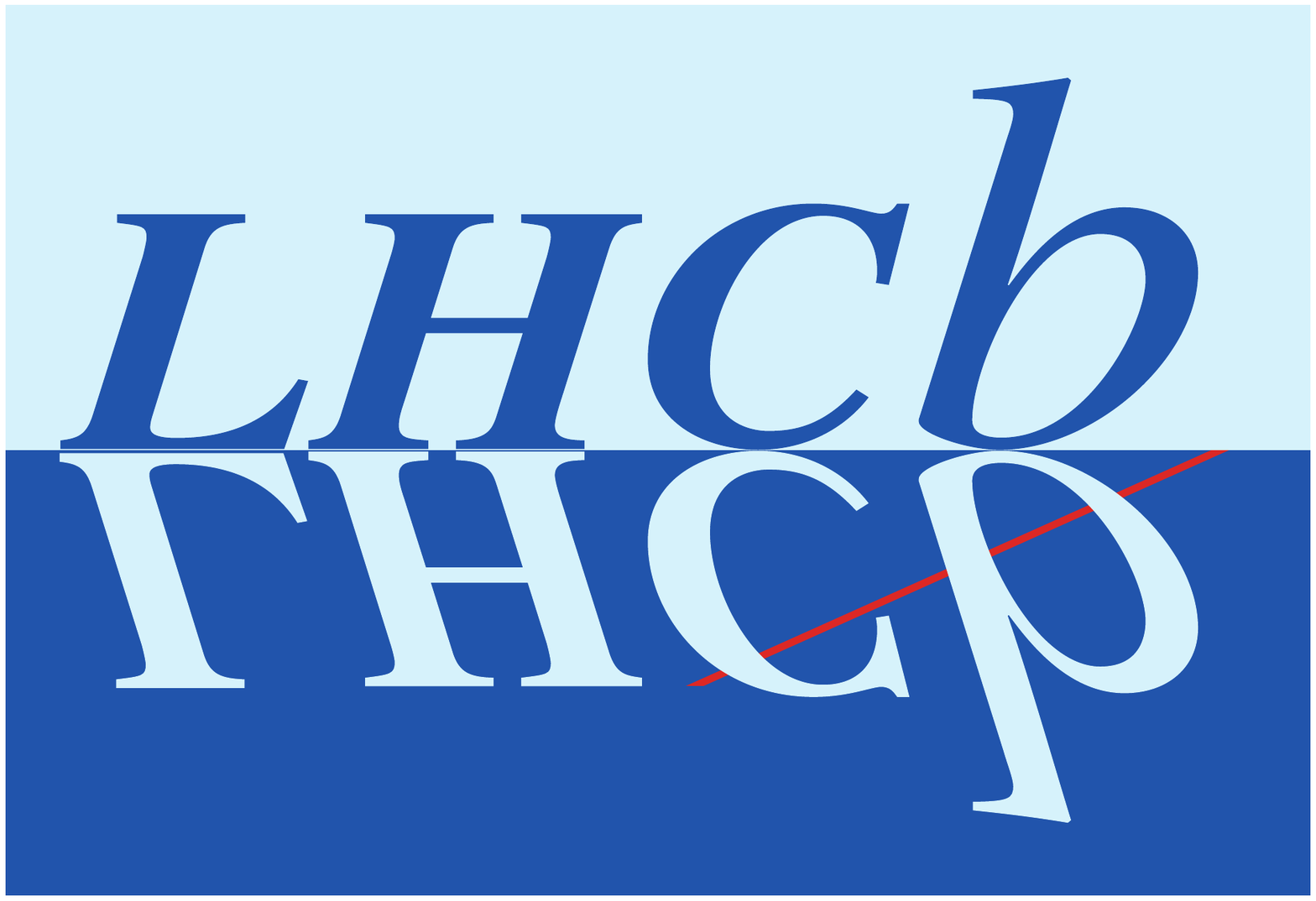}} & &}%
{\vspace*{-1.2cm}\mbox{\!\!\!\includegraphics[width=.12\textwidth]{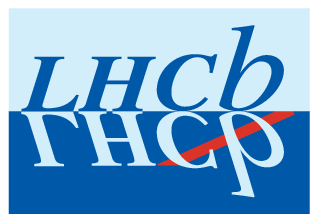}} & &}%
\\
 & & CERN-PH-EP-2015-127 \\  
 & & LHCb-PAPER-2015-020 \\  
 & & May 26, 2015 \\ 
 & & \\
\end{tabular*}

\vspace*{0.5cm}

{\bf\boldmath\huge
\begin{center}
  Study of $\Bm\to D\Km\pip\pim$ and $\Bm\to D\pim\pip\pim$ decays and determination of the CKM angle $\gamma$
\end{center}
}

\vspace*{1.5cm}

\begin{center}
The LHCb collaboration\footnote{Authors are listed at the end of this paper.}
\end{center}

\vspace{\fill}

\begin{abstract}
  \noindent
We report a study of the suppressed $\Bm\to D\Km\pip\pim$ and favored $\Bm\to D\pim\pip\pim$ decays, 
where the neutral $D$ meson is detected through its decays to the $K^{\mp}\pi^{\pm}$ and \CP-even
$\Kp\Km$ and $\pip\pim$ final states. The measurement is carried out using 
a proton-proton collision data sample collected by the LHCb experiment,
corresponding to an integrated luminosity of 3.0\invfb.
We observe the first significant signals in the \CP-even final states of the $D$ meson 
for both the suppressed $\Bm\to D\Km\pip\pim$ and favored $\Bm\to D\pim\pip\pim$ modes, as well as
in the doubly Cabibbo-suppressed $D\to\Kp\pim$ final state of the $\Bm\to D\pim\pip\pim$ decay.
Evidence for the ADS suppressed decay $\Bm\to D\Km\pip\pim$, with $D\to\Kp\pim$, is also presented.
From the observed yields in the $\Bm\to D\Km\pip\pim$, $\Bm\to D\pim\pip\pim$ and their charge conjugate 
decay modes, the most probably value of the weak phase $\gamma$ corresponds to $\gamma=(74^{+20}_{-19})^{\rm o}$. This is 
one of the most precise single-measurement determinations of $\gamma$ to date.
\end{abstract}

\vspace*{1.0cm}

\begin{center}
  Published in Phys.~Rev.~D 
\end{center}

\vspace{\fill}

{\footnotesize 
\centerline{\copyright~CERN on behalf of the \lhcb collaboration, licence \href{http://creativecommons.org/licenses/by/4.0/}{CC-BY-4.0}.}}
\vspace*{2mm}

\end{titlepage}


\newpage
\setcounter{page}{2}
\mbox{~}
%
%
%
%

\cleardoublepage


\renewcommand{\thefootnote}{\arabic{footnote}}
\setcounter{footnote}{0}



\pagestyle{plain} 
\setcounter{page}{1}
\pagenumbering{arabic}


%


\section{Introduction}
The study of beauty and charm hadron decays provides a powerful probe to search for physics
beyond the Standard Model that is complementary to direct searches for new, high-mass particles.
In the Standard Model, the flavor-changing charged currents of quarks are described by the 3$\times$3 unitary 
complex-valued Cabibbo-Kobayashi-Maskawa (CKM) mixing matrix~\cite{Cabibbo:1963yz,Kobayashi:1973fv}, whose elements, 
$V_{\ij}$ ($i=u,c,t$ and $j=d,s,b$), quantify the relative
$i\leftrightarrow j$ coupling strength. Its nine matrix elements can be expressed in terms of four independent
parameters, which need to be experimentally determined. 

In general, decay rates
that involve the $i\leftrightarrow j$ quark transition are sensitive to the magnitudes of the CKM matrix elements, $|V_{ij}|$.
The (weak) phases between different CKM matrix elements can be probed by studying the interference between two (or more) decay
amplitudes. Particle and antiparticle amplitudes are related
by the \CP operator, where $C$ signifies charge conjugation, and $P$ refers to the parity operator. 
Under the \CP operation, weak phases flip sign, leading to different decay rates for particles and antiparticles,
if the weak and (\CP-invariant) strong phases differ between the contributing amplitudes. Precision measurements 
of the magnitudes and phases of the CKM elements provide constraints on many possible scenarios for physics beyond the 
Standard Model.

One of the least well-measured phases is $\gamma\equiv\text{arg}[ - (V_{ud}V^{*}_{ub})/(V_{cd}V^{*}_{cb}) ]$, which can be 
probed by studying the interference between $b\to u$ and $b\to c$ transitions. The most promising method to determine $\gamma$ 
is to study the interference between $\Bm\to\Dz\Km$ and $\Bm\to\Dzb\Km$ decays, when states accessible to both
the $\Dz$ and $\Dzb$ mesons are selected. These modes are 
particularly attractive for the determination of $\gamma$ because their amplitudes are dominated
by only a pair of tree-level processes, leading to a small theoretical uncertainty~\cite{Brod:2013sga}.
Hereafter, we use $D$ without a charge designation when the charm meson can be either a $\Dz$ or $\Dzb$.
A number of methods, depending on the $D$ decay mode, have been discussed in the literature, and are often grouped into
three categories: (i) \CP eigenstates, such as $D\to\Kp\Km$ and $D\to\pip\pim$ decays~\cite{Gronau:1991dp,Gronau:1990ra} (GLW); 
(ii) flavor-specific final states, such as the Cabibbo-favored (CF) and doubly Cabibbo suppressed (DCS) $D\to K^{\pm}\pi^{\mp}$ 
decays~\cite{Atwood:1996ci,Atwood:2000ck} (ADS); 
and (iii) multi-body self-conjugate final states, such as $D\to\KS\pip\pim$~\cite{Giri:2003ty} (GGSZ)\footnote{The letters in the brackets are commonly used to refer to these
general approaches, after the original authors.}. 

Beyond this simplest set of modes, these techniques are also applicable to modes with vector mesons, such as 
$\Bm\to\D^*\Km$, $\Bzb\to D\Kstarzb$~\cite{Dunietz:1991yd}, and $\Bs\to D\phi$~\cite{Gronau:1990ra}, as well
as $b$-baryon decays, e.g. $\Lb\to D\Lz$~\cite{Dunietz:1992ti,FAYYAZUDDIN:1999,Giri:2001ju} decays. It has also been
suggested that other multi-body final states of the recoiling strange quark system could be 
useful~\cite{Gronau:2002mu}, due to the larger branching fractions to these final states, and 
potentially a larger interference contribution. 

The current experimental measurements, averaged over several decays modes, are $\gamma=(73^{+9}_{-10})^{\rm o}$ 
by the LHCb collaboration~\cite{LHCb-PAPER-2013-020,*LHCb-CONF-2014-004},
$\gamma=(69^{+17}_{-16})^{\rm o}$ by the BaBar collaboration~\cite{Lees:2013nha}, and  $\gamma=(68^{+15}_{-14})^{\rm o}$
by the Belle collaboration~\cite{Belle:GammaCombo}. The overall precision on $\gamma$ from a global fit to
direct measurements of $\gamma$ is about 
$7^{\rm o}$~\cite{Charles:2015gya}. To improve the
overall precision on $\gamma$, it is important to study a wide range of final states.

In this article, we present the first ADS and GLW analyses of the decay $\Bm\to DX_s^-$, where 
the $D$ meson is observed through its decay to $K^{\pm}\pi^{\mp}$, $\Kp\Km$ and $\pip\pim$ final states and
$X_s^-\equiv\Km\pip\pim$. When specific charges are indicated in a decay, charge conjugation is implicitly 
included, except in the definition of asymmetries discussed below.
The measurements use proton-proton ($pp$) collision data collected by the LHCb experiment,
corresponding to an integrated luminosity of 3.0\invfb, of which 1.0\invfb was recorded at
a center-of-mass energy of 7\tev and 2.0\invfb at 8\tev. 

\section{Formalism}

The formalism that was developed to describe the $\Bm\to D\Km$ modes can be applied in the
$\Bm\to DX_s^-$ case with only minor modifications~\cite{Gronau:2002mu}.
The decay rates in the \CP final states can be expressed as
\begin{align}
\Gamma (\Bm\to [h^-h^+]_D X_s^-) &\propto 1 + r_B^2 + 2 \kappa r_B \cos(\delta_B - \gamma),\\
\Gamma (\Bp\to [h^-h^+]_D X_s^+) &\propto 1 + r_B^2 + 2 \kappa r_B \cos(\delta_B + \gamma).
\end{align}
Here, $h^{\pm}=\pi^{\pm}$ or $K^{\pm}$, and $[h^-h^+]_D$ indicates that the state in brackets
is produced in the decay of the neutral $D$ meson. The quantities
$r_B\equiv|A(\Bm\to[h^-h^+]_{\Dzb} X_s^-)/A(\Bm\to [h^-h^+]_{\Dz} X_s^-)|$ and $\delta_B$ are the
amplitude ratio and strong phase difference between $\Bm\to\Dzb X_s^-$ and
$\Bm\to\Dz X_s^-$ contributions, averaged over the $DX_s^-$ phase space.
The parameter $\kappa$ is a coherence factor that accounts for a dilution of the
interference due to the variation of the strong phase across the phase space; its value
is bounded between 0 and 1. In principle, $\kappa$ can be obtained in a model-dependent way
by a full amplitude analysis of this decay. Here, we consider it as a free parameter to be
determined in the global fit for $\gamma$. The strong parameters, $r_B$, $\delta_B$ and $\kappa$ are 
specific to this decay, and differ from those obtained from other $B\to DK$ modes. 

The decay rates for the $D\to K^{\pm}\pi^{\mp}$ final states can be written as
\begin{align}
\Gamma (\Bm\to [\Kp\pim]_D X_s^-) &\propto r_B^2 + r_D^2 + 2 \kappa r_Br_D \cos(\delta_B + \delta_D - \gamma), \label{eq:adsmodes1}\\
\Gamma (\Bp\to [\Km\pip]_D X_s^+) &\propto r_B^2 + r_D^2 + 2 \kappa r_Br_D \cos(\delta_B + \delta_D + \gamma),\label{eq:adsmodes2}\\
\Gamma (\Bm\to [\Km\pip]_D X_s^-) &\propto 1 + (r_Br_D)^2 + 2 \kappa r_Br_D \cos(\delta_B - \delta_D - \gamma),\\
\Gamma (\Bp\to [\Kp\pim]_D X_s^+) &\propto 1 + (r_Br_D)^2 + 2 \kappa r_Br_D \cos(\delta_B - \delta_D + \gamma).
\label{eq:adsmodes}
\end{align}
Here, additional parameters $r_D$ and $\delta_D$ enter, which quantify the ratio of the
DCS to CF amplitude, $A(\Dz\to\Kp\pim)/A(\Dz\to\Km\pip)=r_De^{i\delta_D}$. 
Values of $r_D$ and $\delta_D$ are taken from independent measurements~\cite{HFAG,Ablikim:2014gvw}.

The determination of the \CP observables in the $\Bm\to DX_s^-$ decay uses the favored $\Bm\to D\pim\pip\pim$ decay for normalization,
denoted here as $\Bm\to D X_d^-$. For brevity, we will use $X^-$ to refer to either $X_d^-$ or $X_s^-$.
In addition, $D\to K\pi$ is used when both charge combinations are considered.

The \CP observables of interest for the GLW analysis are the charge-averaged yield ratios
\begin{align}
R^{h^+h^-}_{\CP+}& \equiv 2\frac{\Gamma(\Bm\to [h^+h^-]_D X_s^-) + \Gamma(\Bp\to[h^+h^-]_DX_s^+)}{ \Gamma(\Bm\to [\Km\pip]_DX_s^-) + \Gamma(\Bp\to [\Kp\pim]_DX_s^+) } \nonumber \\
  & = 1 + r_B^2 + 2\kappa r_B \cos\delta_B\cos\gamma. 
\label{eq:rglw}
\end{align}
\noindent 
Because of the different $D$ final states in Eq.~\ref{eq:rglw}, systematic uncertainty due to
the precision of the $D$ branching fractions and the different selections is incurred. Following
Ref.~\cite{Gronau:2002mu}, we neglect \CP violation in the $\Bm\to DX_d^-$ and the favored $D$ final state 
of $\Bm\to DX_s^-$ decays, and approximate $R^{h^+h^-}_{\CP+}$ by the following double ratio
\begin{align}
R_{\CP+} \simeq \frac{R^{h^+h^-}_{s/d}}{R^{K\pi}_{s/d}},
\label{eq:rcpapprox}
\end{align}
\noindent where
\begin{align}
R^{h^+h^-}_{s/d}&\equiv \frac{\Gamma(\Bm\to [h^+h^-]_D X_s^-) + \Gamma(\Bp\to [h^+h^-]_DX_s^+)}{\Gamma(\Bm\to [h^+h^-]_DX_d^-) + \Gamma(\Bp\to [h^+h^-]_DX_d^+) },\label{eq:rglwapp1} \\
R^{K\pi}_{s/d}&\equiv \frac{\Gamma(\Bm\to [\Km\pip]_D X_s^- + \Gamma(\Bp\to [\Kp\pim]_D X_s^+)}{\Gamma(\Bm\to [\Km\pip]_DX_d^- + \Gamma(\Bp\to [\Kp\pim]_D X_d^+)}. 
\label{eq:rglwapp2}
\end{align}
This double ratio has the benefit that almost all systematic uncertainties cancel to first order.
The neglected \CP-violating contribution of magnitude $\kappa r_B|\Vus\Vcd/\Vud\Vcs|\lesssim0.01$
is included as a source of systematic uncertainty. 

We also make use of the charge asymmetries
\begin{align}
\Asy^{f}_{X^{\pm}}\equiv \frac{\Gamma(\Bm\to f_DX^-) - \Gamma(\Bp\to\bar{f}_D X^+)}{ \Gamma(\Bm\to f_DX^-) + \Gamma(\Bp\to\bar{f}_DX^+) } = 2\kappa r_B\sin\delta_B\sin\gamma / R_{\CP+},
\label{eq:aglw}
\end{align}
\noindent where $f$ refers to either $\Kp\Km$, $\pip\pim$ or the CF $\Km\pip$ final state in the $D$ meson decay. For
simplicity, small contributions from direct \CP violation in $D\to\pip\pim$ and $D\to\Kp\Km$ are not included here, but
are accounted for in the fit for $\gamma$~\cite{LHCb-PAPER-2013-020,*LHCb-CONF-2014-004}. 

For the ADS modes, we measure the relative widths of the DCS to CF decays, separated by charge, as
\begin{align}
R^{X^{\pm}} = \frac{\Gamma(B^{\pm}\to [K^{\mp}\pi^{\pm}]_DX^{\pm})}{ \Gamma(B^{\pm}\to [K^{\pm}\pi^{\mp}]_DX^{\pm})} 
= \frac{r_B^2+r_D^2+2\kappa r_Br_D\cos(\delta_B+\delta_D\pm\gamma)}{1+r_B^2r_D^2+2\kappa r_B r_D\cos(\delta_B-\delta_D\pm\gamma)}.
\label{eq:rads}
\end{align}
\noindent The nearly identical final states in these ratios lead to a cancellation of the most significant
sources of systematic uncertainty. Corrections to $R^{X^{\pm}}$ for $\Dz-\Dzb$ mixing~\cite{Rama:2013voa} are omitted for clarity,
but are included in the fit for $\gamma$~\cite{LHCb-PAPER-2013-020,*LHCb-CONF-2014-004}. 

All of the above equations, except for Eqs.~\ref{eq:rcpapprox}--\ref{eq:rglwapp2}, can be applied to
either $B^{\pm}\to D X_s^{\pm}$ or $B^{\pm}\to D X_d^{\pm}$ decays. The values of 
$r_B$, $\delta_B$ and $\kappa$ differ between the favored and suppressed decays; however $\gamma$ is common to both.
Most of the sensitivity is expected to come from the $B^{\pm}\to D X_s^{\pm}$ decays, 
since $A(\Bm\to\Dzb X_d^-)/A(\Bm\to\Dz X_d^-)$ is ${\mathcal{O}}(\lambda^2)$, as compared to ${\mathcal{O}}(1)$ for 
$A(\Bm\to\Dzb X_s^-)/A(\Bm\to\Dz X_s^-)$,  where $\lambda=0.2253\pm0.0014$~\cite{PDG2014} is the sine of the Cabibbo angle.
Taken together, the observables that contain the most significant information on $\gamma$ are 
$R_{\CP+}$, $\Asy^{h^+h^-}_{X_s}$ and $R^{X_s^{\pm}}$. Measurements of these four quantities constrain
$r_B$, $\delta_B$, $\kappa$ and $\gamma$.

The product branching fraction for $\Bm\to D X_s^-$ decays, with $D\to h^+h^-$, is at the level of about $10^{-6}$. The small branching fractions,
combined with a total selection efficiency that is of order 0.1\%, makes the detection and study of these modes challenging.
The corresponding ADS DCS decay mode is expected to have a yield of at least 10 times less than the \CP modes, and is very sensitive to 
the values of $r_B$, $\delta_B$, $\kappa$, and $\gamma$ (see Eqs.~\ref{eq:adsmodes1} and~\ref{eq:adsmodes2}). For this reason, the signal region of the ADS suppressed
decays (both $\Bm\to DX_d^-$ and $\Bm\to DX_s^-$) was not examined until all selection requirements were determined.

\section{The LHCb detector and simulation}
The \lhcb detector~\cite{Alves:2008zz} is a single-arm forward
spectrometer covering the \mbox{pseudorapidity} range $2<\eta <5$,
designed for the study of particles containing \bquark or \cquark
quarks. The detector includes a high-precision tracking system
consisting of a silicon-strip vertex detector surrounding the $pp$
interaction region, a large-area silicon-strip detector located
upstream of a dipole magnet with a bending power of about
$4{\rm\,Tm}$, and three stations of silicon-strip detectors and straw
drift tubes~\cite{LHCb-DP-2013-003} placed downstream of the magnet.
The combined tracking system provides a momentum measurement with
a relative uncertainty that varies from 0.5\% at low momentum, \ptot, to 1.0\% at 200\gevc,
and an impact parameter measurement with a resolution of about 20\mum~\cite{LHCb-DP-2014-001}  for
charged particles with large transverse momentum, \pt. The polarity of the dipole magnet is
reversed periodically throughout data-taking to reduce asymmetries in the detection of charged particles.
Different types of charged hadrons are distinguished using information
from two ring-imaging Cherenkov detectors~\cite{LHCb-DP-2012-003}. Photon, electron and
hadron candidates are identified by a calorimeter system consisting of
scintillating-pad and preshower detectors, an electromagnetic
calorimeter and a hadronic calorimeter. Muons are identified by a
system composed of alternating layers of iron and multiwire
proportional chambers~\cite{LHCb-DP-2012-002}. Details on
the performance of the LHCb detector can be found in Ref.~\cite{LHCb-DP-2014-002}.

The trigger~\cite{LHCb-DP-2012-004} consists of a
hardware stage, based on information from the calorimeter and muon
systems, followed by a software stage, which applies a full event
reconstruction. The software trigger requires a two-, three- or four-track
secondary vertex with a large \pt sum of the 
tracks and a significant displacement from all primary $pp$
interaction vertices~(PVs). At least one particle should have 
$\pt > 1.7\gevc$ and $\chisqip$ with respect to any PV
greater than 16, where $\chisqip$ is defined as the
difference in \chisq of a given PV reconstructed with and
without the considered particle. A multivariate algorithm~\cite{BBDT} is used for
the identification of secondary vertices consistent with the decay
of a \bquark-hadron. 

Proton-proton collisions are simulated using
\pythia~\cite{Sjostrand:2006za,*Sjostrand:2007gs} with a specific \lhcb
configuration~\cite{LHCb-PROC-2010-056}.  Decays of hadronic particles
are described by \evtgen~\cite{Lange:2001uf}, in which final-state
radiation is generated using \photos~\cite{Golonka:2005pn}. The
interaction of the generated particles with the detector, and its
response, are implemented using the \geant toolkit~\cite{Allison:2006ve, *Agostinelli:2002hh} as described in
Ref.~\cite{LHCb-PROC-2011-006}. In modeling the $\Bm\to DX^-$ decays, we include 
several resonant and nonresonant contributions to emulate the $X_s^-$ and $X_d^-$ systems, as well
as contributions from orbitally excited $D$ states, e.g $D_1(2420)^{0}\to\Dz\pip\pim$. The contributions
are set based on known branching fractions, or tuned to reproduce resonant substructures seen in the data.

\section{Candidate selection}

Candidate $\Bm$ decays are reconstructed by combining a $D\to K\pi$, $D\to\Kp\Km$ or $D\to\pip\pim$ 
candidate with an $X^-$ candidate. A kinematic fit~\cite{Hulsbergen:2005pu} is performed, where several 
constraints are imposed: the reconstructed positions of the
$X^-$ and $\Bm$ decay vertices are required to be compatible with each other,
the $D$ candidate must point back to the $\Bm$ decay vertex, the $\Bm$ candidate must have a direction
consistent with originating from a PV in the event, and the invariant mass of the $D$ candidate must be consistent with
the known $\Dz$ mass~\cite{PDG2014}.
The production point of each $\Bm$ candidate is designated to be the PV for which the $\chisqip$ is smallest.

Candidate $D$ mesons are required to have invariant mass within $3\sigma_D$ ($2.5\sigma_D$ for $D\to\pim\pip$ decays) of the known value,
where the mass resolution, $\sigma_D$, varies from 7.0\mevcc for $D\to\Kp\Km$ to 10.2\mevcc for $D\to\pip\pim$ decays. Unlike the $D$ mesons, the
invariant mass of the $X^-$ system covers a broad range from about $0.9-3.3$\gevcc. Candidates are 
required to have an invariant mass, $M(X^-)<2.0$\gevcc. For the $X_s^-$ system, we also require the $\Km\pip$ invariant mass to be within 100\mevcc of the known 
$\Kstarz$ mass. 
The latter two requirements not only improve the signal-to-background ratio, but should also increase the coherence factor 
$\kappa$ in the final state.

To improve the signal-to-background ratio further, we select candidates based on particle identification (PID) information, and on the output 
of a boosted decision tree (BDT)~\cite{Breiman, AdaBoost} classifier. The latter discriminates signal from combinatorial background based on 
information derived primarily from the tracking system. For the BDT, signal efficiencies are obtained from large
samples of simulated signal decays. Particle identification efficiencies are obtained from a large $\Dstarp\to\Dz\pip$ calibration data 
sample~\cite{LHCb-DP-2012-003}, reweighted in $\pt$, $\eta$ and number of tracks in
the event to match the distributions in data. The effect of the BDT and PID selection requirements
on the background is assessed using sidebands well away from the $\Bm$ peak region. In the optimization, a wide range of selection
requirements on the PID and BDT outputs are scanned, and we choose the value that optimizes the expected statistical precision 
of the $\Bm\to DX_s^-$ signal yield. Expected signal yields are evaluated based on known or estimated branching fractions and efficiencies obtained from 
simulation (for the BDT) or $\Dstarp\to\Dz\pip,~\Dz\to\Km\pip$ calibration data (for the PID).
Due to the smaller expected yields in the ADS modes, separate optimizations are performed for the GLW and the ADS analyses.
Using simulated decays, we find that the relative efficiencies for $\Bm\to DX_s^-$ and $\Bm\to DX_d^-$ decays across the
phase space are compatible for the GLW and ADS selections. Due to the uniformity of the selections,
and the fact that the observables are either double ratios, e.g. $R_{\CP+}$, or
ratios involving almost identical final states, the systematic uncertainty on the relative efficiencies is negligible compared to the statistical uncertainty.

Several other mode-specific requirements are imposed to suppress background from other $b$-hadron decays.
First, we explicitly veto contributions from 
$\Bm\to\Dz\Dsm$, with either $\Dsm\to\pim\pip\pim$ or $\Dsm\to\Km\pip\pim$, by rejecting candidates in which the $X^-$ system has 
invariant mass within 15\mevcc of
the known $\Dsm$ mass. Contamination from other final states that include a charmed particle are also sought by forming all 
two-, three- and four-body
combinations (except the $D\to h^+h^{\prime-}$ signal decay), and checking for peaks at any of the known
charmed particle masses. Contributions from $\Dz\to\Km\pip,~\Km\pip\pim\pip$, $\Ds\to\Kp\Km\pip$ and $\Dp\to\Km\pip\pip$ decays are seen,
and $\pm$15\mevcc mass vetoes are applied around the known charm particle masses. In addition,
$\Dstarp$ contributions are removed by requiring the invariant mass difference, $M[(\Km\pip)_D\pip]-M[(\Km\pip)_D]>148.5$\mevcc. This removes both 
partially reconstructed $B\to\Dstarp X$ final states and fully reconstructed states, such as
$\Bm\to D_1(2420)^{0}h^-$, $D_1(2420)^{0}\to\Dstarp\pim$, $\Dstarp\to\Dz\pip$ signal decays. The latter, while forming a good signal candidate,
are flavor-specific, and therefore would reduce the coherence of the final state.
Those $\D^{**0}\to\Dz\pip\pim$  contributions that do not have a $\Dstarp$ intermediate state are kept, since they are not flavor-specific.

Another potentially large source of background is from five-body charmless $B$ decays. Unfortunately, their branching fractions are generally unknown, 
but they are likely to be sizable compared to those of the $\Bm\to DX_s^-$ signal decays. Moreover, these backgrounds could have large \CP asymmetries, as
seen in three-body $B$-meson decays~\cite{LHCb-PAPER-2014-044,LHCb-PAPER-2014-034,PDG2014}. It is therefore important 
to suppress their contribution to a negligible level. This is investigated by applying all of the above selections, except 
that $D$ candidates are selected from a $D$ mass sideband region instead of the signal region. The sideband region is chosen to avoid the contribution
from the other two-body $D$ decays with one misidentified daughter. Charmless backgrounds are seen in all modes. 
These backgrounds are reduced to a negligible level by requiring that the $D$ decay vertex is displaced significantly downstream of the
$\Bm$ decay vertex, corresponding to three times the uncertainty on the measured $D$ decay length. A more stringent requirement,
corresponding to five times the uncertainty on the measured $D$ decay length, is imposed on the $\Bm\to [\pip\pim]_DX_{s,d}^-$ decays,
which is found to have a much larger charmless contribution. After these requirements are applied, the charmless backgrounds are consistent 
with zero, and the residual contribution is considered as a source of systematic uncertainty.

Another important background to suppress is the cross-feed from the ADS CF $\Bm\to [\Km\pip]_DX^-$ decay into the 
ADS DCS $\Bm\to [\Kp\pim]_DX^-$ sample, which may happen if the $\Km$ and $\pip$ are both misidentified.
Since the CF yield is expected to be several hundred times larger than that of the DCS mode 
(depending on the values of $r_B$, $\delta_B$, $\kappa$ and $\gamma$), a large suppression is necessary.
The combined $\Dz$ mass and PID requirements provide a suppression factor of $6\times10^{-5}$. 
An additional requirement that the $K\pi$ invariant mass (after interchanging the $\Km$ and $\pip$ masses) differs by at least 15\mevcc 
from the known $\Dz$ mass decreases the suppression level to $0.9\times10^{-5}$. This leads to a negligible
contamination from the CF ADS mode into the DCS decay. The same veto is applied to both the ADS CF $\Dz\to\Km\pip$
and DCS $\Dz\to\Kp\pim$ decays, so that no efficiency correction is needed for $R^{X^{\pm}}$.

Lastly, in order to have a robust estimate of the trigger efficiency for signal events, we impose requirements on information from the hardware 
trigger; either (i) one or more of the decay products of the signal candidate met the trigger requirements from the calorimeter system, or
(ii) the event passed at least one of the hardware triggers, and would have done so even if the signal decay was removed from the event.
These two classes of events constitute about 60\% and 40\% of the signal candidates, respectively, where the overlap
is assigned to category (i).

\begin{figure}
    \centering
    \includegraphics[width=0.98\textwidth]{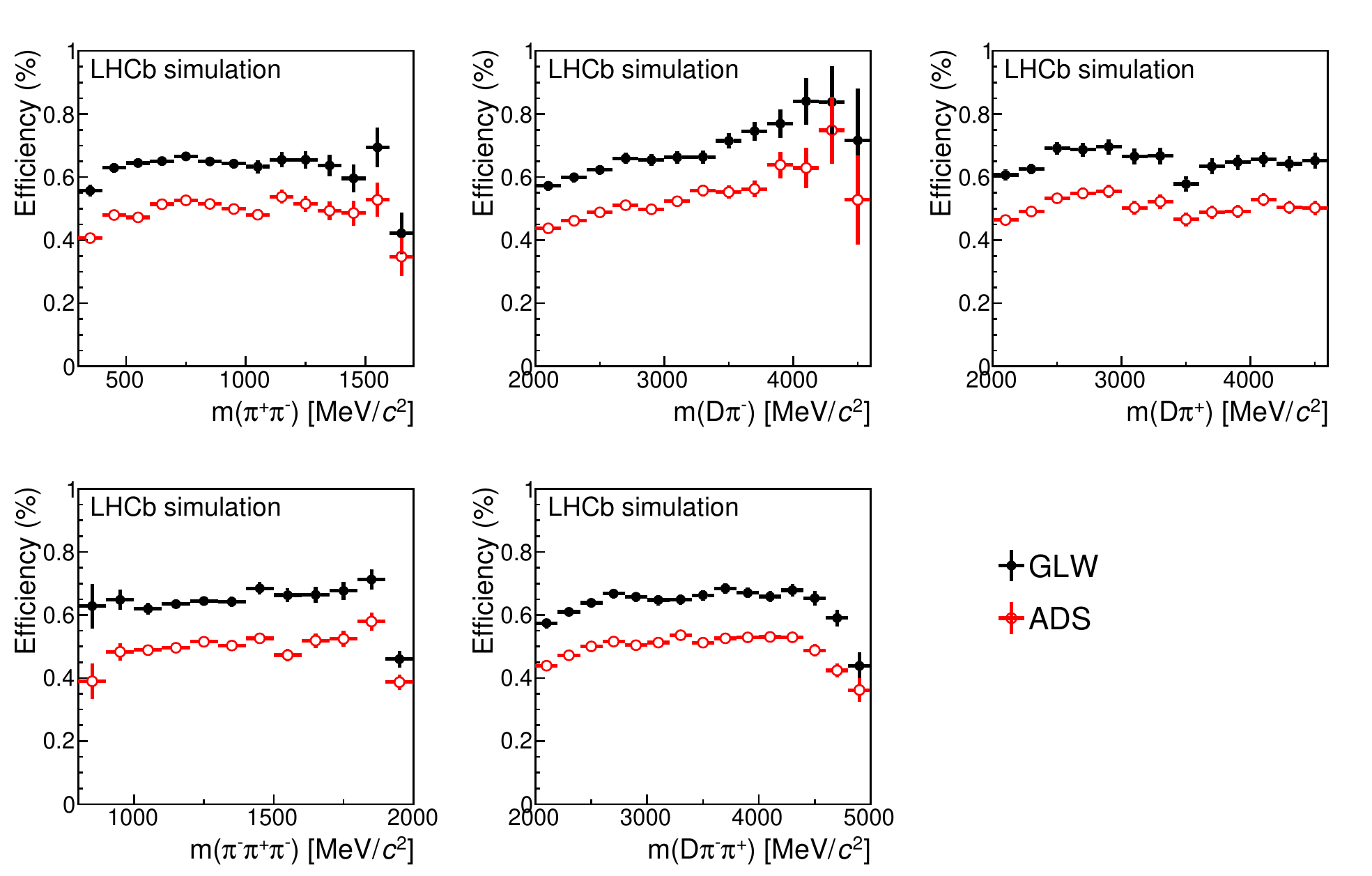}
    \caption{\small{Signal efficiencies for the $\Bm\to [\Km\pip]_DX_d^-$ decay when applying the GLW and
ADS selections. The efficiencies are shown as a function of five different two- and three-body masses.}}
    \label{fig:EffOverlayDalitz_B2D0A1_Paper}
\end{figure}

\begin{figure}
    \centering
    \includegraphics[width=0.98\textwidth]{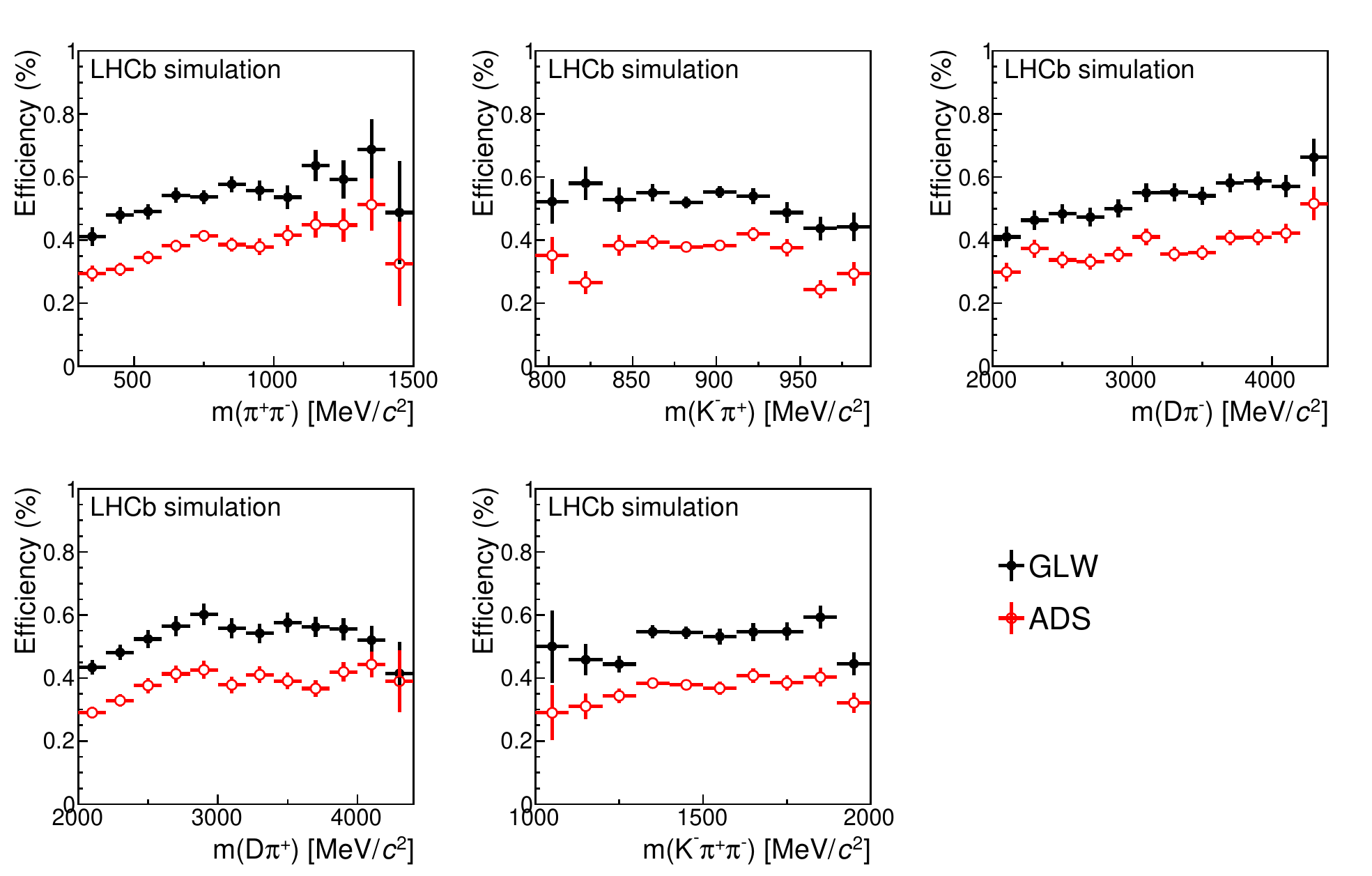}
    \caption{\small{Signal efficiencies for the $\Bm\to [\Km\pip]_DX_s^-$ decay when applying the GLW and
ADS selections. The efficiencies are shown as a function of five different two- and three-body masses.}}
    \label{fig:EffOverlayDalitz_B2D0K1_Paper}
\end{figure}

The selection efficiencies as a function of several two- and three-body masses in the $\Bm\to [\Km\pip]_DX_d^-$ decay are
shown in Fig.~\ref{fig:EffOverlayDalitz_B2D0A1_Paper}, for both the GLW and ADS selections. The 
efficiencies for other $D$ final states are consistent with those for $\D\to\Km\pip$. The $m(D\pim)$ and
$m(\pip\pim)$ efficiencies include two entries per signal decay, as there are two $\pim$ in the final state.
The analogous efficiencies for the $\Bm\to DX_s^-$ decay are shown in Fig.~\ref{fig:EffOverlayDalitz_B2D0K1_Paper}. The
relative efficiencies of the ADS to GLW selections are consistent with being flat across each of these masses. 
These efficiencies include all selection requirements, including PID.
However, events in which any of the signal 
decay products is outside of the LHCb detector acceptance are not included, since they are not simulated; 
thus to obtain the total selection efficiency, these efficiencies should be scaled by a factor of 0.11, as
determined from simulation.

\begin{figure}
    \centering
    \includegraphics[width=0.48\textwidth]{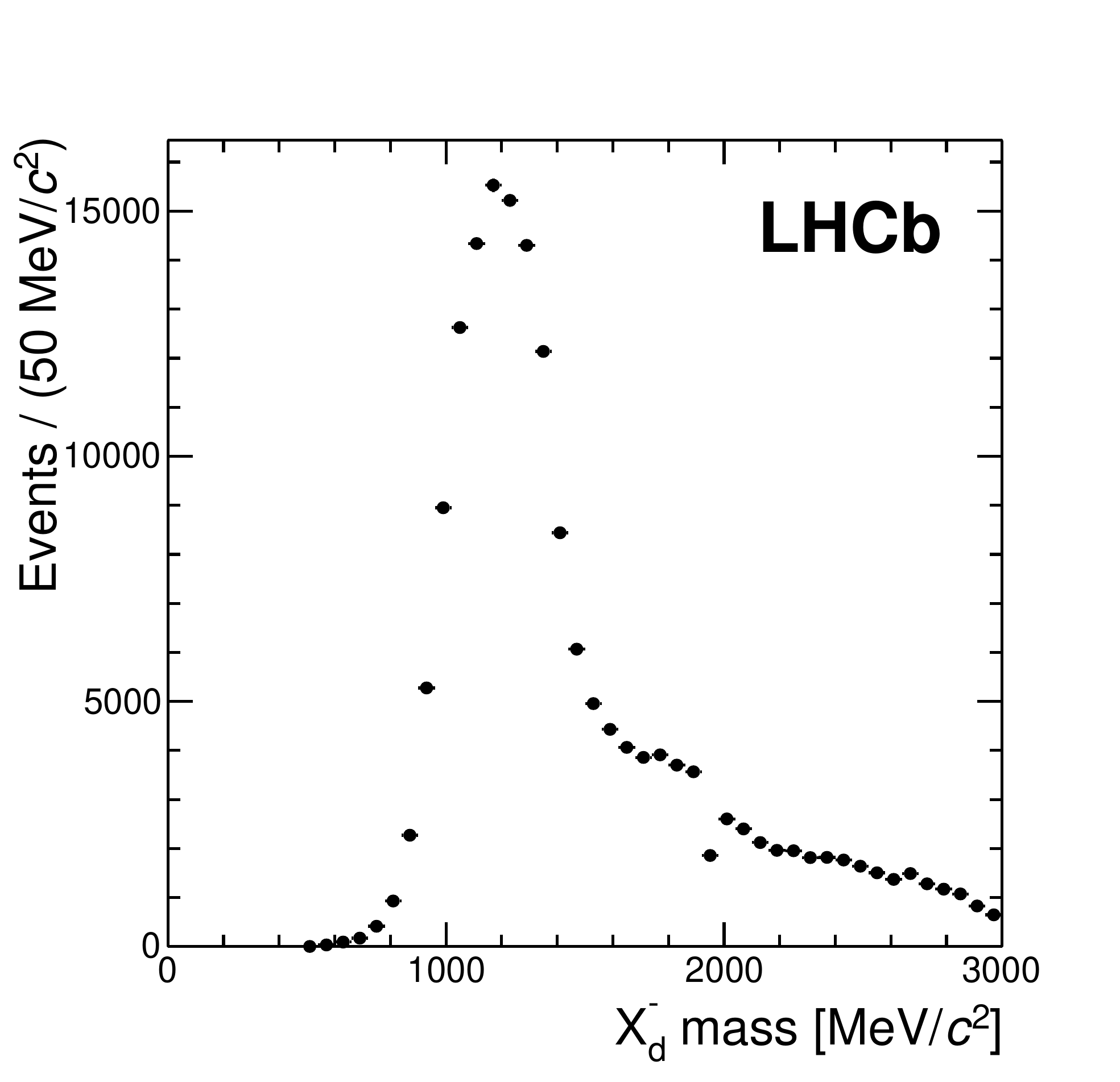}
    \includegraphics[width=0.48\textwidth]{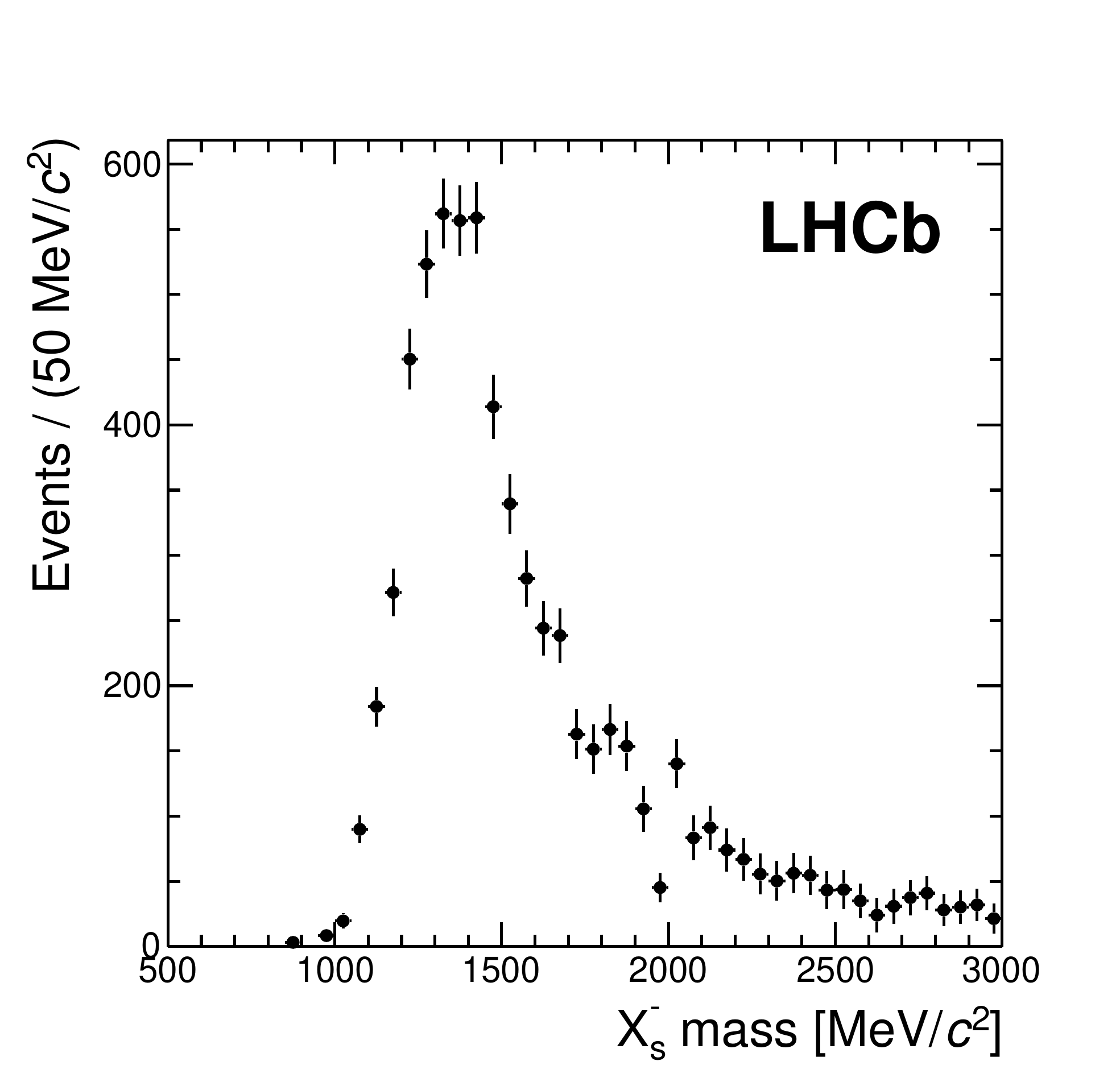}
    \caption{\small{Signal distributions of the (left) $X_d^-$ invariant mass in $\Bm\to DX_d^-$ decays and (right) $X_s^-$ invariant mass, 
      in $\Bm\to DX_s^-$ decays, for $D\to\Km\pip$. The distributions are obtained using the \sPlot method. In both cases, all selections, 
      except the $M(X^-)<2$\gevcc and the $\Kstarz$ mass selection, are applied. The dip at 1.97\gevcc is due to the $\Dsp$ meson veto.}}
    \label{fig:Xmass}
\end{figure}

Figure~\ref{fig:Xmass} shows the $X_d^-$ and $X_s^-$ invariant mass distributions for $\Bm\to [\Km\pip]_DX_d^-$ and $\Bm\to [\Km\pip]_DX_s^-$
signal decays after all selections, except for the $X^-$ and $\Kstarz$ mass requirements. These signal spectra are background subtracted
using the \sPlot method~\cite{Pivk:2004ty}, with the $\Bm$ candidate invariant mass as the discriminating variable.
The $X_d^-$ and $X_s^-$ contributions peak in the region below 2\gevcc, consistent with the dominance of resonances such as 
$a_1(1260)^-\to\pim\pip\pim$
to the $X_d^-$ system, and one or more excited strange resonances contributing to $X_s^-$. The dip at 1.97\gevcc is due to the $\Dsm$ mass veto. 

\section{Fits to data}

The signal yields are determined through a simultaneous unbinned extended maximum likelihood fit to the 
16 $B^{\pm}$ candidate invariant mass spectra. These
16 spectra include the four $\Bm\to DX_d^-$ decays, where $D\to K^{\pm}\pi^{\mp},~\Kp\Km$ and $\pip\pim$, the 
corresponding four charge-conjugate decays, and the set of eight modes where $X_d^-$ is replaced with $X_s^-$. 
The signal and background contributions across these modes are similar, although not identical. Where possible, common
signal and background shapes are used; otherwise simulation is used to relate parameters in the lower yield modes to 
the values obtained from the high yield CF $D\to K\pi$ modes. Signal and background yields are all independent
of one another in the $\Bp$ and $\Bm$ mass fits; thus \CP violation is allowed for all contributions in the mass spectrum. 
Unless otherwise noted, the shapes discussed below are obtained from simulated decays. 

\subsection{Signal shapes}

The $\Bm$ mass signal shapes are each parameterized as the sum of a Crystal Ball (CB) shape~\cite{Skwarnicki:1986xj} and a Gaussian ($G$) function,
\begin{align}
{\cal{F}}_{sig} \propto f_{\rm CB}{\rm CB}(m_B,\sigma_{\rm CB},\alpha_{\rm CB},n) + (1 - f_{\rm CB}) G(m_B,\sigma_g). 
\end{align} 
\noindent The Gaussian function accounts for the core of the mass distribution, whereas the
CB function accounts for the non-Gaussian radiative tail below, and a wider Gaussian resolution component above, the 
signal peak. A small difference is seen between the shapes for the $\Bm\to DX_d^-$ and $\Bm\to DX_s^-$ decays, and
so a different set of signal shape parameters is used to describe each,
except for a common value of the fitted $\Bm$ mass, $m_B$. The signal shapes are not very sensitive to the
power-law exponent, $n$, which is fixed to 10. The parameters
$\alpha_{\rm CB}$, $\sigma_g$ and $f_{\rm CB}$ are allowed to vary freely in the fit to the data.
From simulation, we find that for all 16 modes, $\sigma_{\rm CB}/\sigma_{g}$ is consistent with 1.90, and this ratio is imposed in the fit. 
Simulation is also used
to relate the mass resolution in the $D\to\Kp\Km,~\pip\pim$ modes to that of the $D\to K\pi$ mode, from which it is found that
$\sigma_g^{[KK]_DX^-} = (0.947\pm0.011)\sigma_g^{[K\pi]_DX^-}$ and $\sigma_g^{[\pi\pi]_DX^-} = (1.043\pm0.011)\sigma_g^{[K\pi]_DX^-}$. 
The relations are consistent between the $\Bm\to DX_d^-$ and $\Bm\to DX_s^-$ modes, and are applied as fixed constraints (without uncertainties)
in the mass fit. 

\subsection{Backgrounds and their modeling}

The primary sources of background in the mass spectra are partially reconstructed $B\to D^{(*)}X^-$ decays, cross-feed between 
$\Bm\to DX_d^-$ and $\Bm\to DX_s^-$, and other combinatorial backgrounds. All of the spectra have a contribution from
combinatorial background, the shape of which is described by an exponential function. Its slope is taken to be the same for 
the \CP-conjugate $\Bm$ and $\Bp$ decays, but differs among the various $D$ and $X^-$ final states. 

The main contribution to the partially reconstructed background comes from
$\Bm\to [\Dz\piz,\Dz\gamma]_{\Dstarz}X^-$ or $\Bzb\to [\Dz\pip]_{\Dstarp}X^-$ decays, where
a pion or photon is not considered when reconstructing the $\Bm$ candidate.
Because the missed pion or photon generally has low momentum, these decays pass the full selection with high efficiency. 
The shapes of these distributions are modeled using parameterized shapes based on simulated decays.
Since the Dalitz structure of these backgrounds is not known, we do not
rely entirely on simulation to reproduce the shape of this low-mass component. Instead,
the parameters of the shape function that depend on the decay dynamics are allowed to vary freely, 
and are determined in the fit. The shape parameters for these backgrounds are varied independently for 
$\Bm\to DX_d^-$ and $\Bm\to DX_s^-$ decays.

Another background contribution which primarily contributes to the $\Bm\to DX_d^-$ ADS suppressed mode
is the $\Bzb\to\Dz\pim\pip\pim\pip$ decay, where there is no $\Dstarp$ intermediate state. 
This decay can contribute to the ADS CF mode if a $\pip$ is excluded from the decay,
or to the ADS DCS mode if a $\pim$ is not considered.
The branching fraction for this decay is not known, but 
the similar CF decay $\Bzb\to\Dstarz\pim\pip\pim\pip$ is known to have a relatively large branching fraction of 
$(2.7\pm0.5)\times10^{-3}$~\cite{Majumder:2004su,Edwards:2001mw}.
Assuming $\brf(\Bzb\to\Dz\pim\pip\pim\pip)\simeq\brf(\Bzb\to\Dstarz\pim\pip\pim\pip)$,
this background contribution is about two orders of magnitude larger than the DCS signal, although it
peaks at lower mass than the signal. The selection efficiency and shape of this background are difficult 
to determine from simulation, since there have not been any studies of this final state to date. Its shape is
obtained from simulations that assume a quasi two-body process, $\Bzb\to\Dz R,~R\to\pim\pip\pim\pip$,
which decays uniformly in the phase space. 
An ARGUS shape~\cite{Albrecht:1990am} convolved with a Gaussian function provides a good description of this
simulated background. Its shape parameters are shared between $\Bp$ and $\Bm$ and are allowed to vary freely in the fit, except for the Gaussian width, 
which is fixed to the expected mass resolution of 15\mevcc. 

The analogous $\Bzb\to\Dz\Km\pip\pim\pip$ decay does not pose the same contamination to the DCS ADS 
$\Bp\to[\Km\pip]_DX_s^+$ signal, since a missed $\pim$ leads to
a $\Bp\to\Dz\Km\pip\pip$ candidate, which is not one of the decays of interest. However, in the 
$\Bsb\to [\Km\pip]_{D^0}\Kp\pim\pip\pim$ decay, opposite-sign kaons are natural due to the presence of the 
$\bar{s}$ quark within the $\Bsb$ meson. This decay is unobserved, but the similar decay, $\Bsb\to\Dz\Kp\pim$,
has a relatively large branching fraction of $(1.00\pm0.15)\times10^{-3}$~\cite{LHCb-PAPER-2014-036}.
Based on other $B$-meson decays, one would expect the $\Bsb\to\Dz\Kp\pim\pip\pim$ decay to be at the same level, ${\cal{O}}(10^{-3})$,
which is two orders of magnitude larger than the signal. The shape of this background has a similar threshold behavior as 
for the $\Bzb\to\Dz\pim\pip\pim\pip$ decay discussed previously, and therefore its contribution is also modeled from simulated decays
using an ARGUS shape convolved with a Gaussian function with freely varying shape parameters. 

In the fit, we also model cross-feed between the $\Bm\to D^{(*)}X_d^-$ and $\Bm\to DX_s^-$ decays.
The shapes of these cross-feed backgrounds are obtained from simulation.
The cross-feed rate is obtained from $\Dstarp\to\Dz\pip$,~$\Dz\to\Km\pip$ calibration data, reweighted to match the
properties of the signal decays. All selection requirements on the $\Bm\to DX^-$ decays, including 
$|M(\Km\pip)-M_{\Kstarz}|<100$\mevcc and $M(X^-)<2$\gevcc, are taken into account.
In total, we find that 0.66\% of $\Bm\to DX_d^-$ are misidentified as  $\Bm\to DX_s^-$ for the GLW modes and
0.16\% for the ADS modes. The lower value for the ADS modes is due to the tighter PID requirements on the $\Km$ candidate
in the $X_s^-$ system. The cross-feed from $\Bm\to DX_s^-$ into $\Bm\to DX_d^-$ is evaluated in an analogous manner,
and is found to be 13.7\%. Since the ratio of
branching fractions is $\brf(\Bm\to DX_s^-)/ \brf(\Bm\to DX_d^-)\simeq 0.09$~\cite{Aaij:2012bw}, the yield of this 
background is only about 1\% of the signal yield.

Other sources of background that contribute to the $\Bm\to DX_s^-$ modes are the $\Bm\to\Dz[\Km\Kp\pim]_{\Dsm}$ and
$\Bm\to\Dz\Km [\Kp\pim]_{\Kstarzb}$ decays, where the $\Kp$ is misidentified as a $\pip$ meson.  The shapes are similar for these two 
backgrounds and thus a single shape is used, based on a parameterization of the $\Bm$ candidate mass distribution in simulated 
$\Bm\to\Dz [\Km\Kp\pim]_{\Dsm}$ decays.
Taking into account known branching fractions~\cite{PDG2014}, efficiencies from simulation, and $\Kp\to\pip$ misidentification rates 
from $\Dstarp\to\Dz\pip$ calibration data, we expect a contribution of 1.6\% of the $\Bm\to DX_s^-$ signal.

\subsection{Fit results}

\begin{figure}
    \centering
    \includegraphics[width=0.48\textwidth]{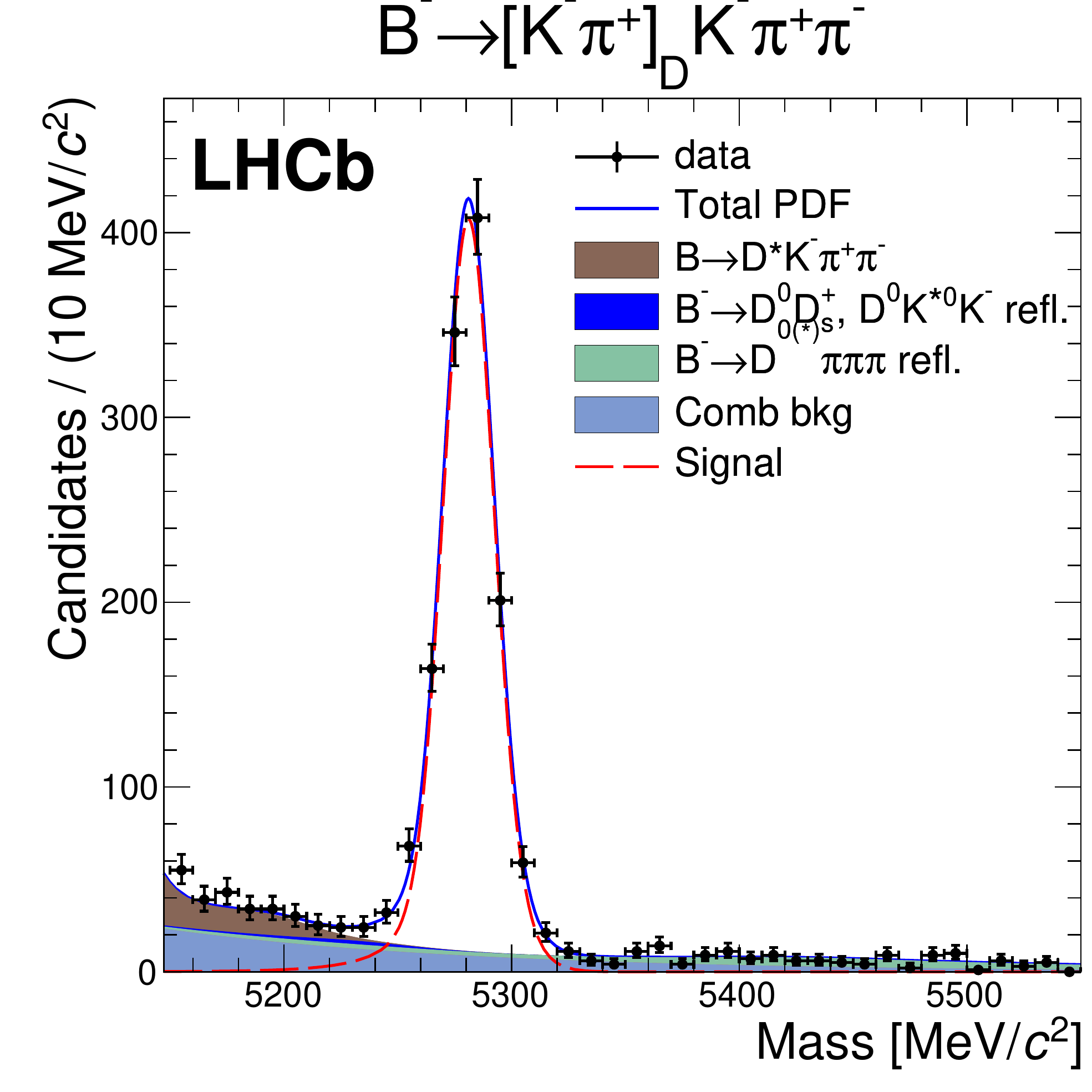}
    \includegraphics[width=0.48\textwidth]{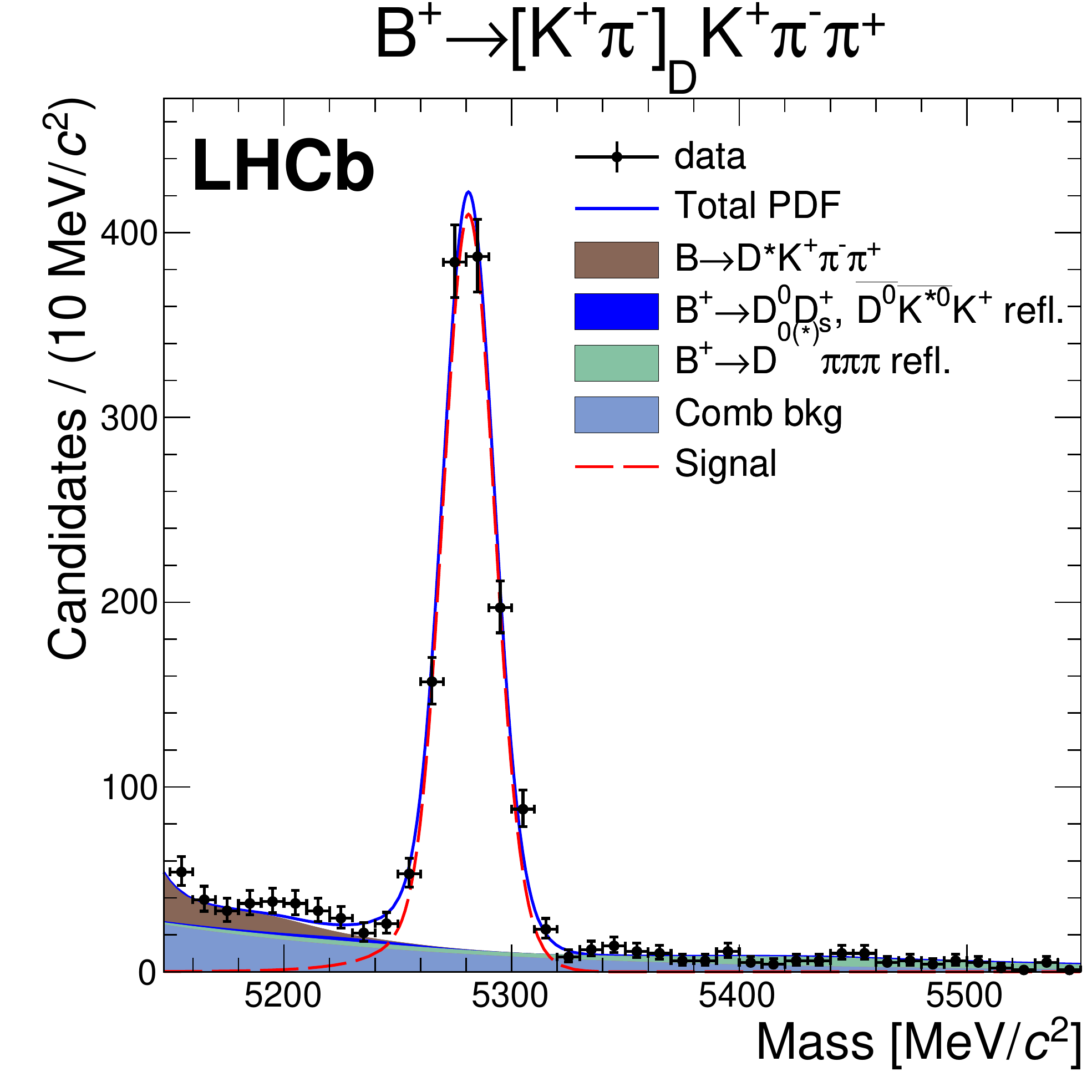}
    \includegraphics[width=0.48\textwidth]{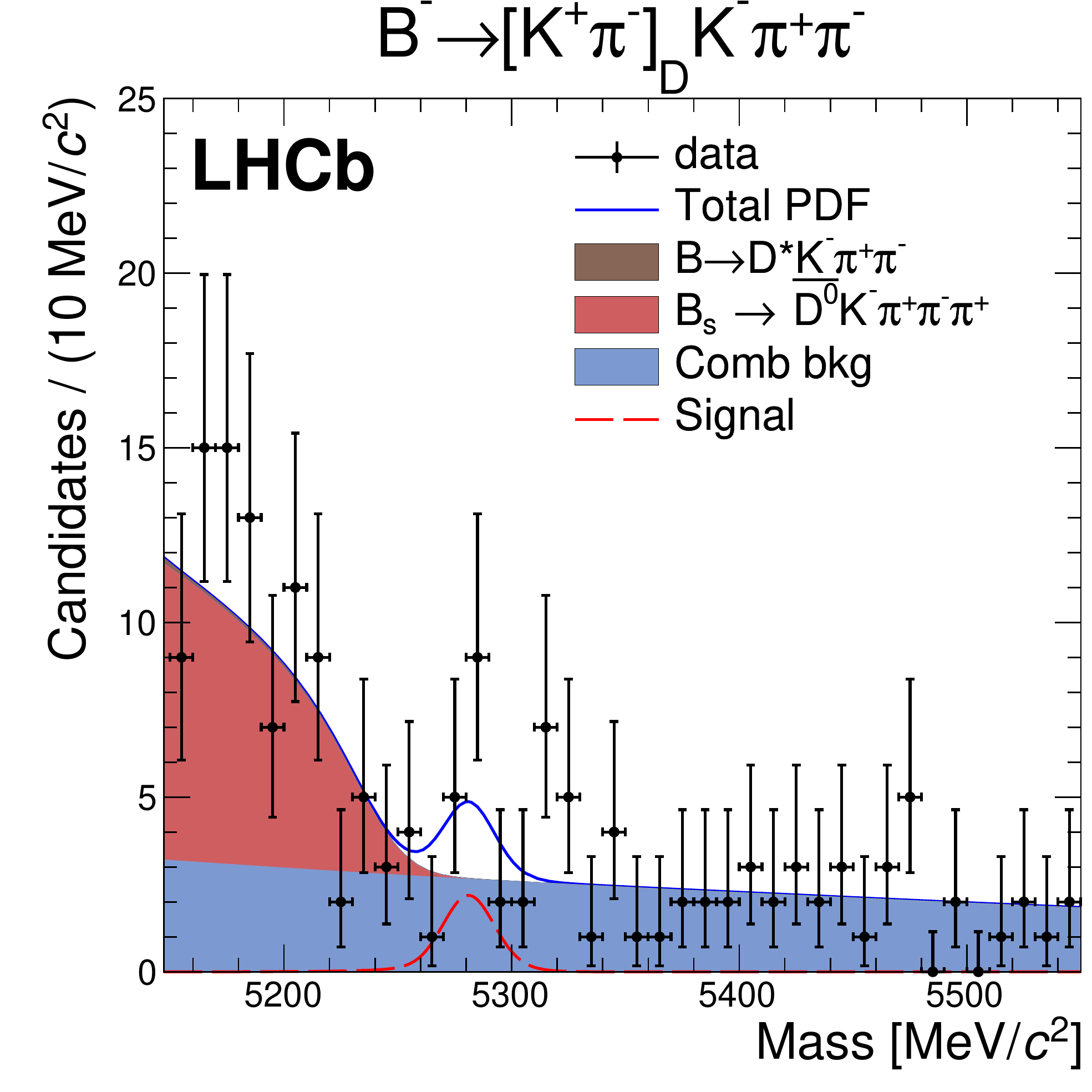}
    \includegraphics[width=0.48\textwidth]{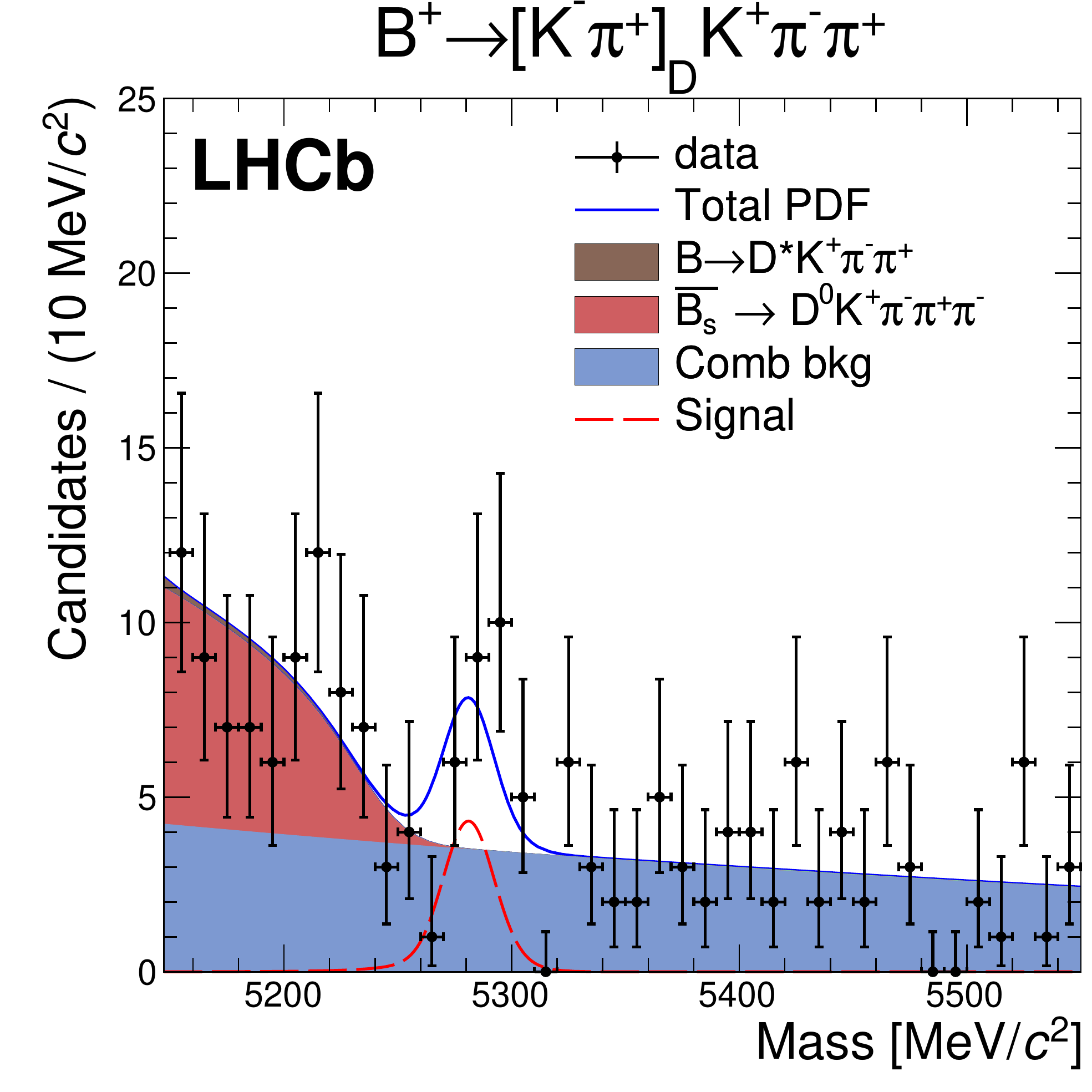}
    \caption{\small{Mass distributions of $\Bm\to DX_s^-$ candidates using the ADS selections, for (top left) $\Bm\to [\Km\pip]_DX_s^-$,
(top right) $\Bp\to [\Kp\pim]_DX_s^+$, (bottom left) $\Bm\to [\Kp\pim]_DX_s^-$, and (bottom right) $\Bp\to [\Km\pip]_DX_s^+$.}}
\label{fig:adsfits_cs}
\end{figure}

\begin{figure}
    \centering
    \includegraphics[width=0.42\textwidth]{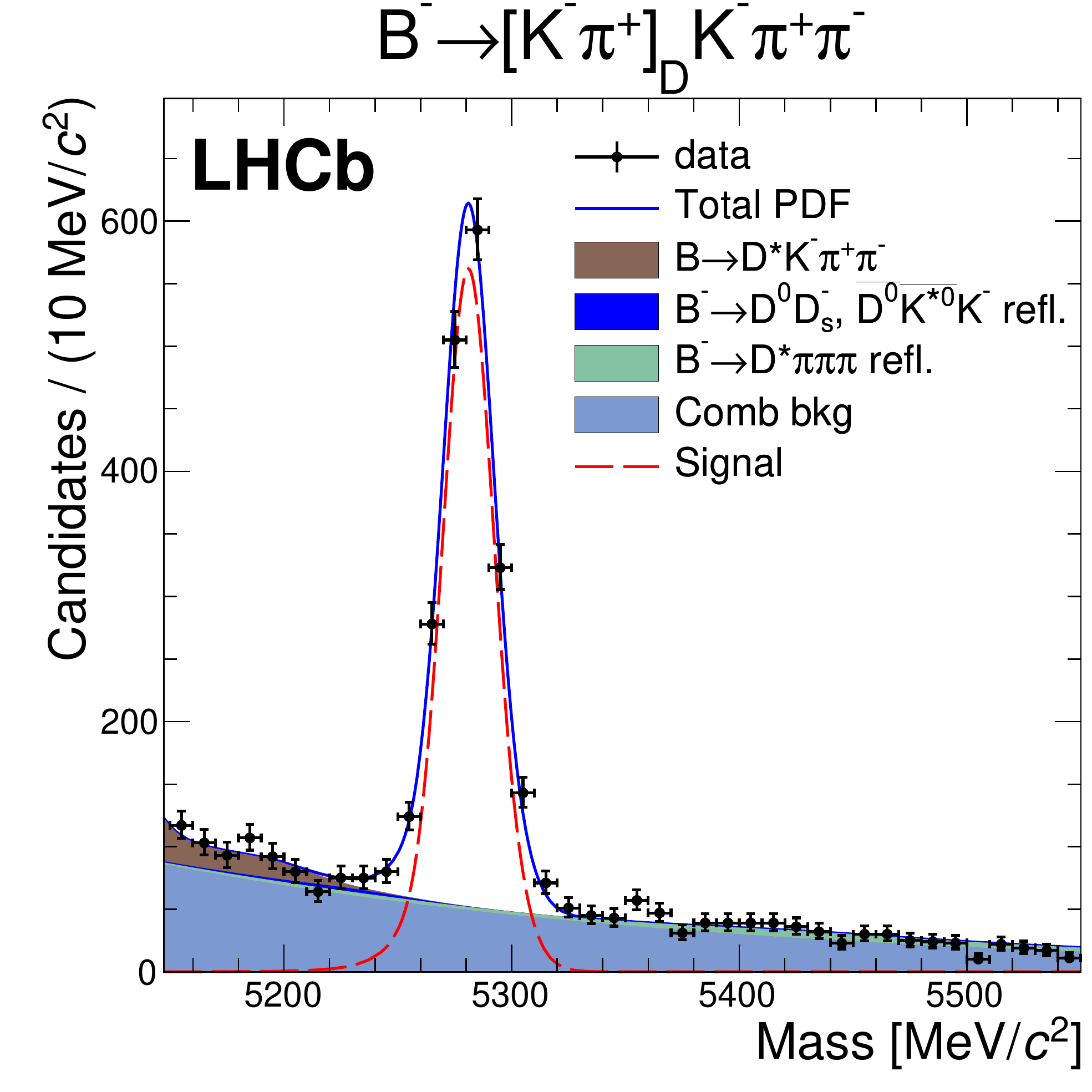}
    \includegraphics[width=0.42\textwidth]{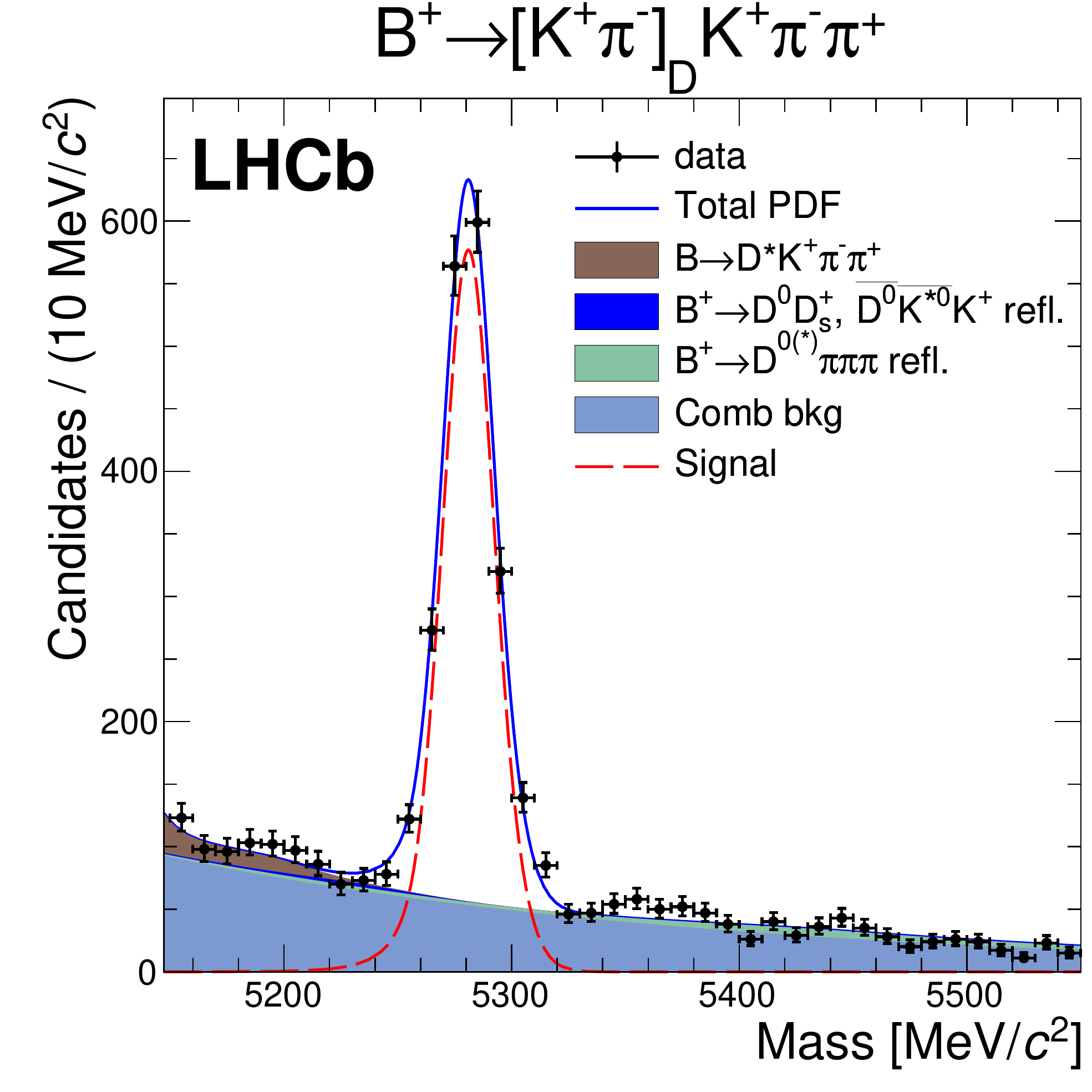}
    \includegraphics[width=0.42\textwidth]{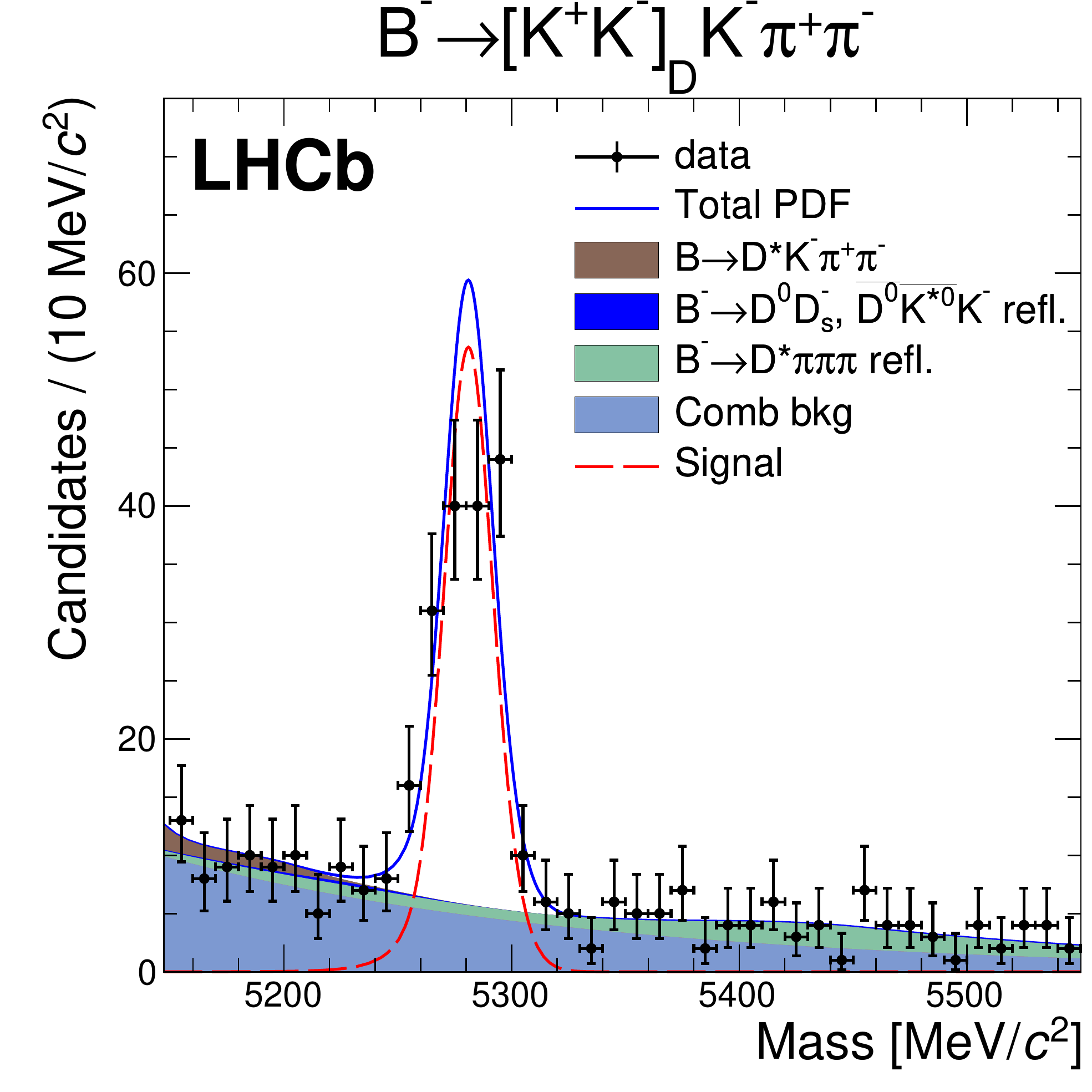}
    \includegraphics[width=0.42\textwidth]{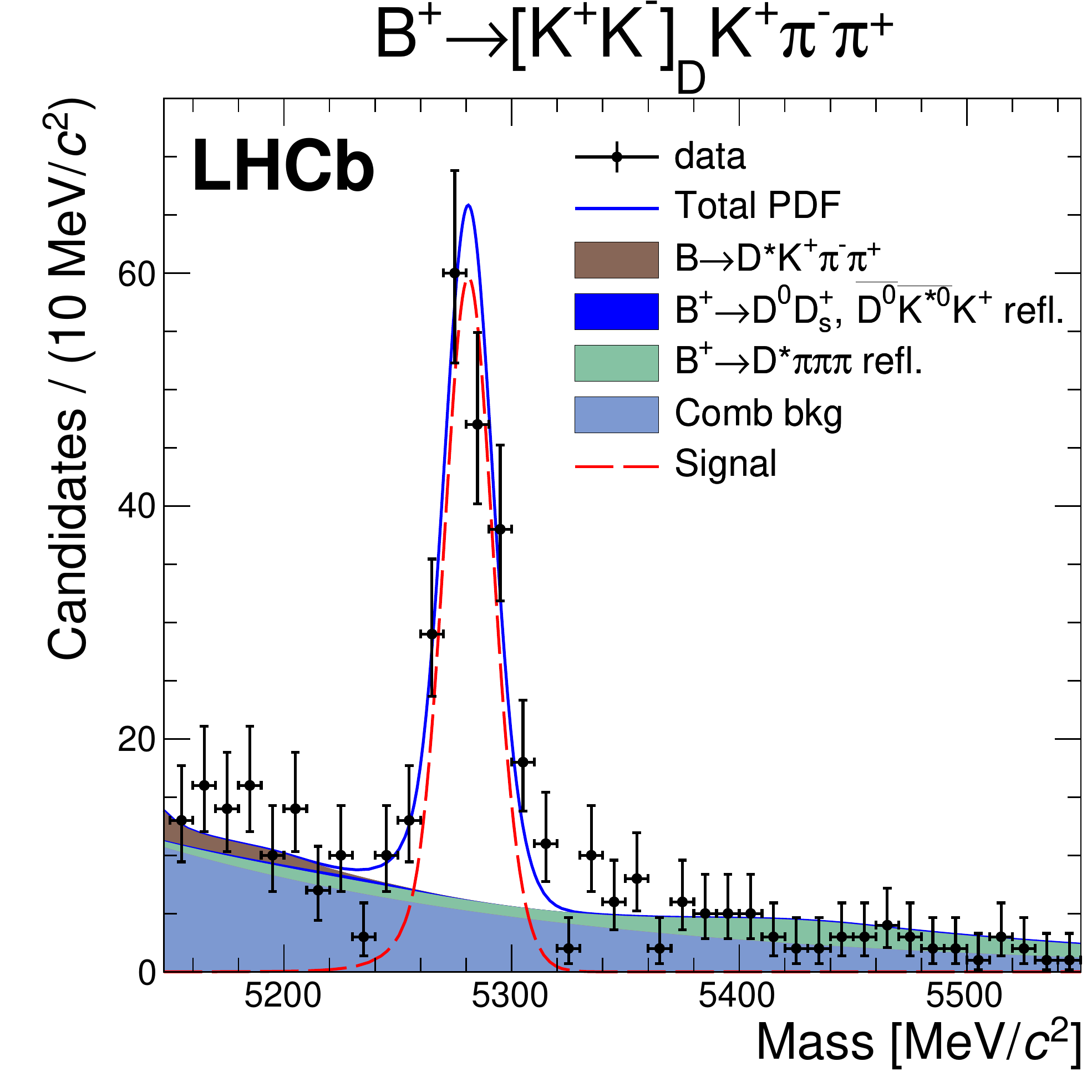}
    \includegraphics[width=0.42\textwidth]{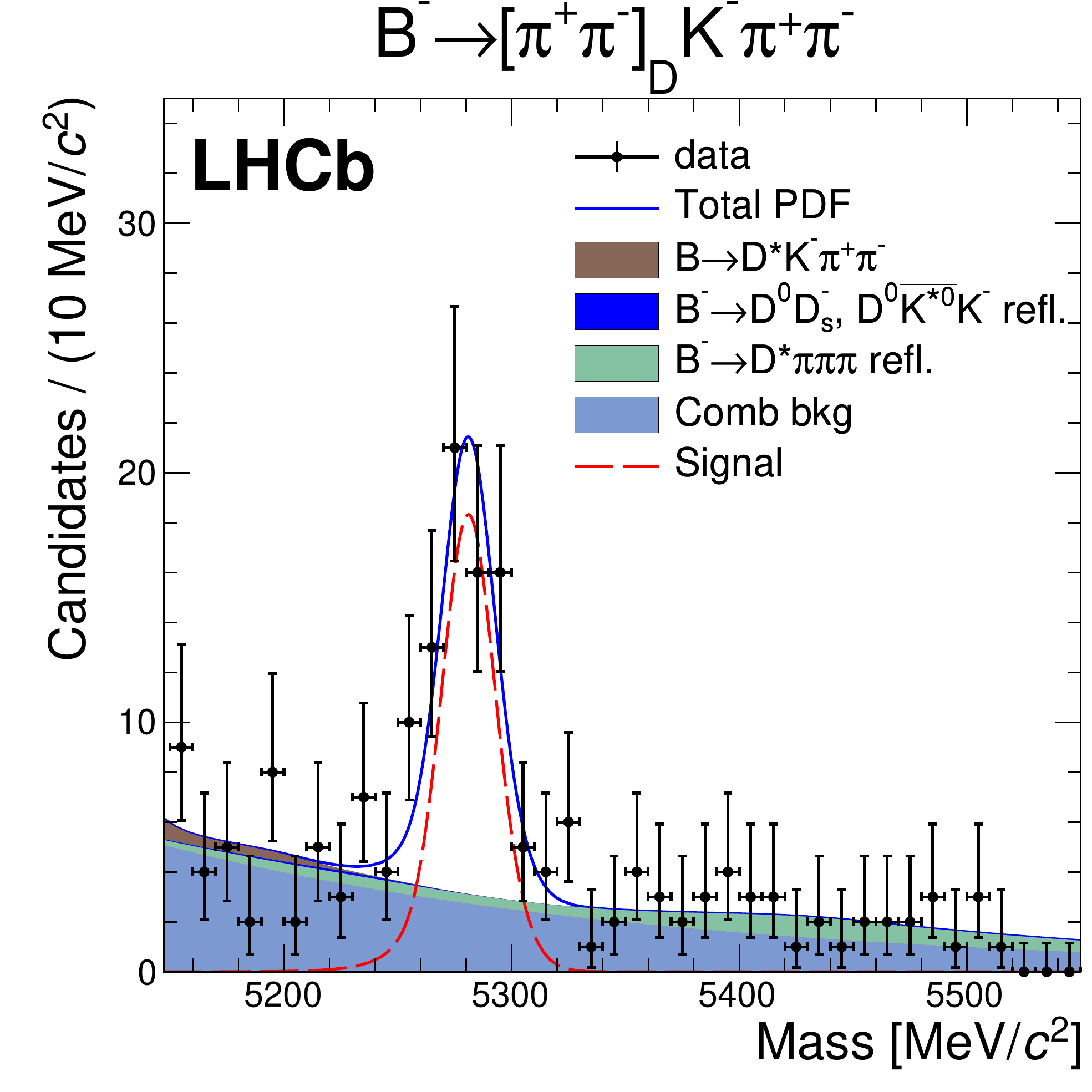}
    \includegraphics[width=0.42\textwidth]{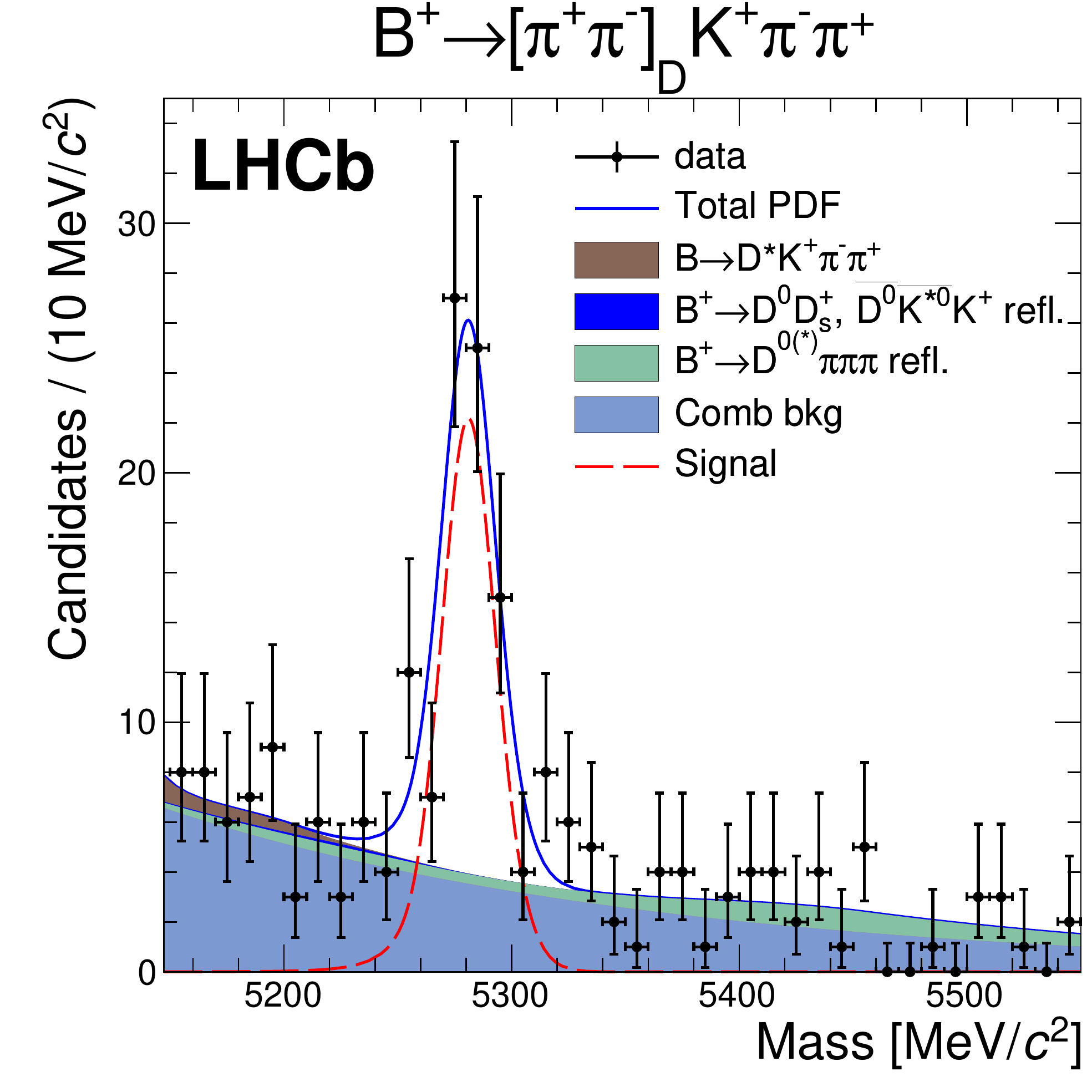}
    \caption{\small{Mass distributions of $\Bm\to DX_s^-$ candidates using the GLW selections, for (top left) $\Bm\to [\Km\pip]_DX_s^-$,
      (top right) $\Bp\to [\Kp\pim]_DX_s^+$, (middle left) $\Bm\to [\Kp\Km]_DX_s^-$, (middle right) $\Bp\to [\Kp\Km]_DX_s^+$, 
        (bottom left) $\Bm\to [\pip\pim]_DX_s^-$, and 
      (bottom right) $\Bp\to [\pip\pim]_DX_s^+$.}}
\label{fig:glwfits_cs}
\end{figure}

\begin{figure}
    \centering
    \includegraphics[width=0.48\textwidth]{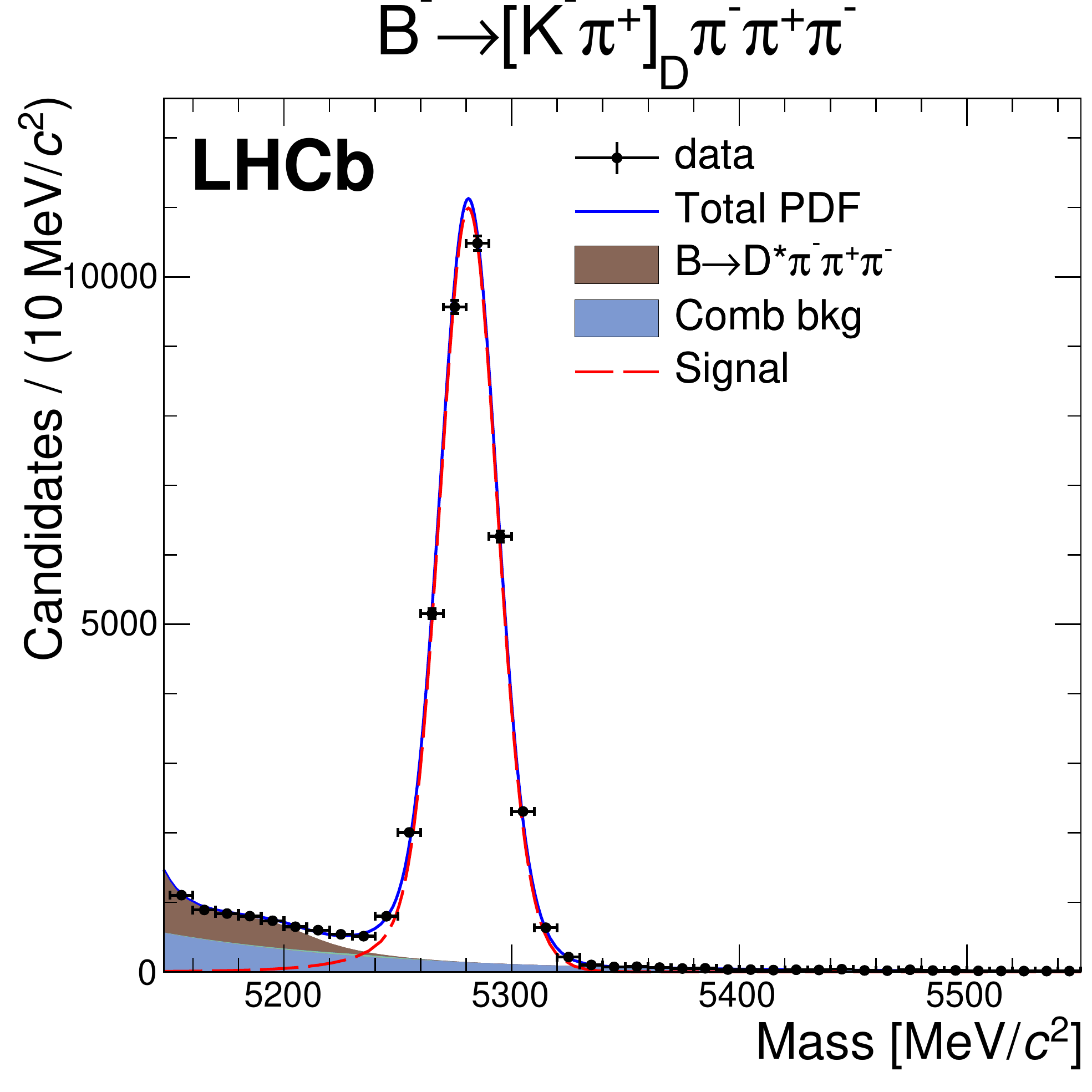}
    \includegraphics[width=0.48\textwidth]{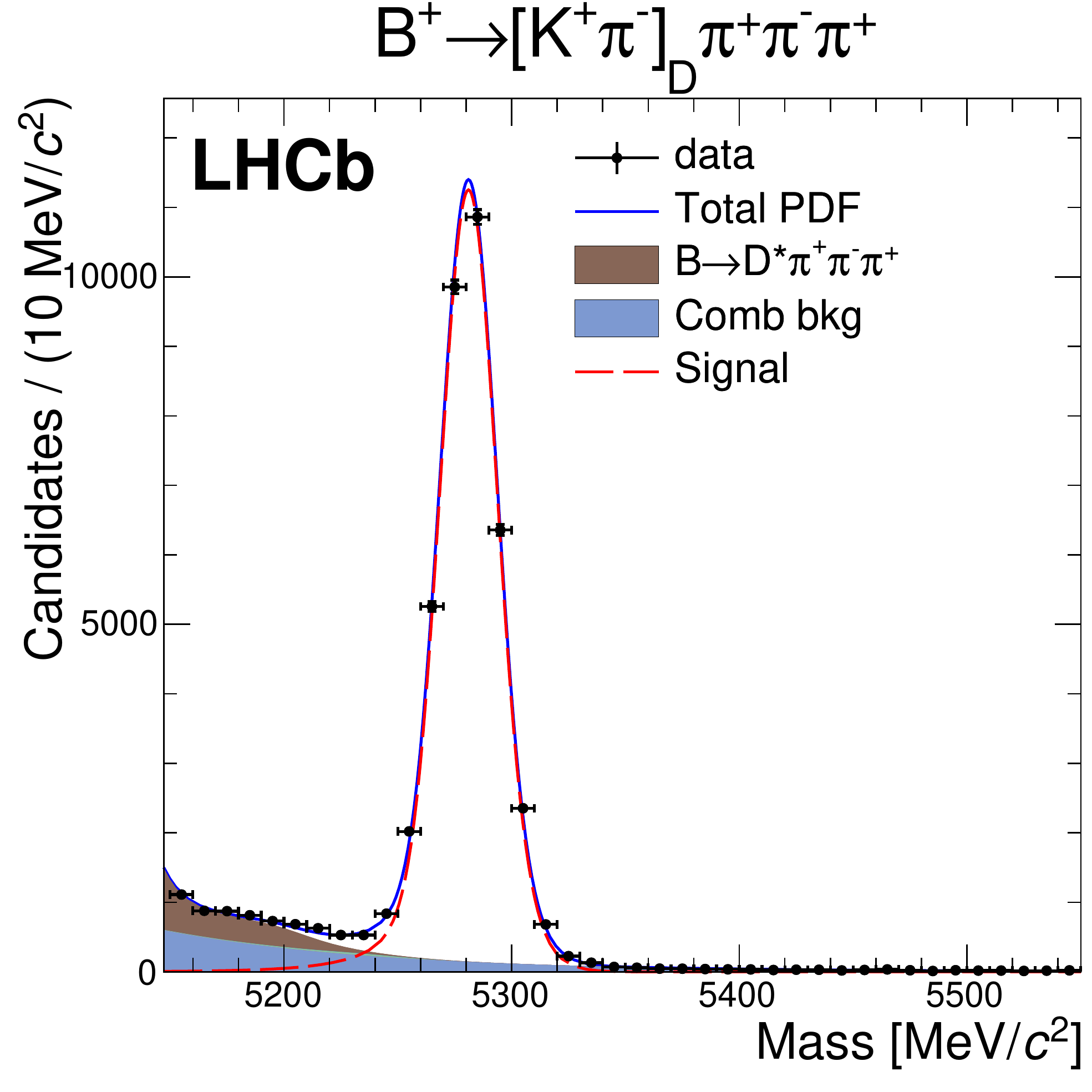}
    \includegraphics[width=0.48\textwidth]{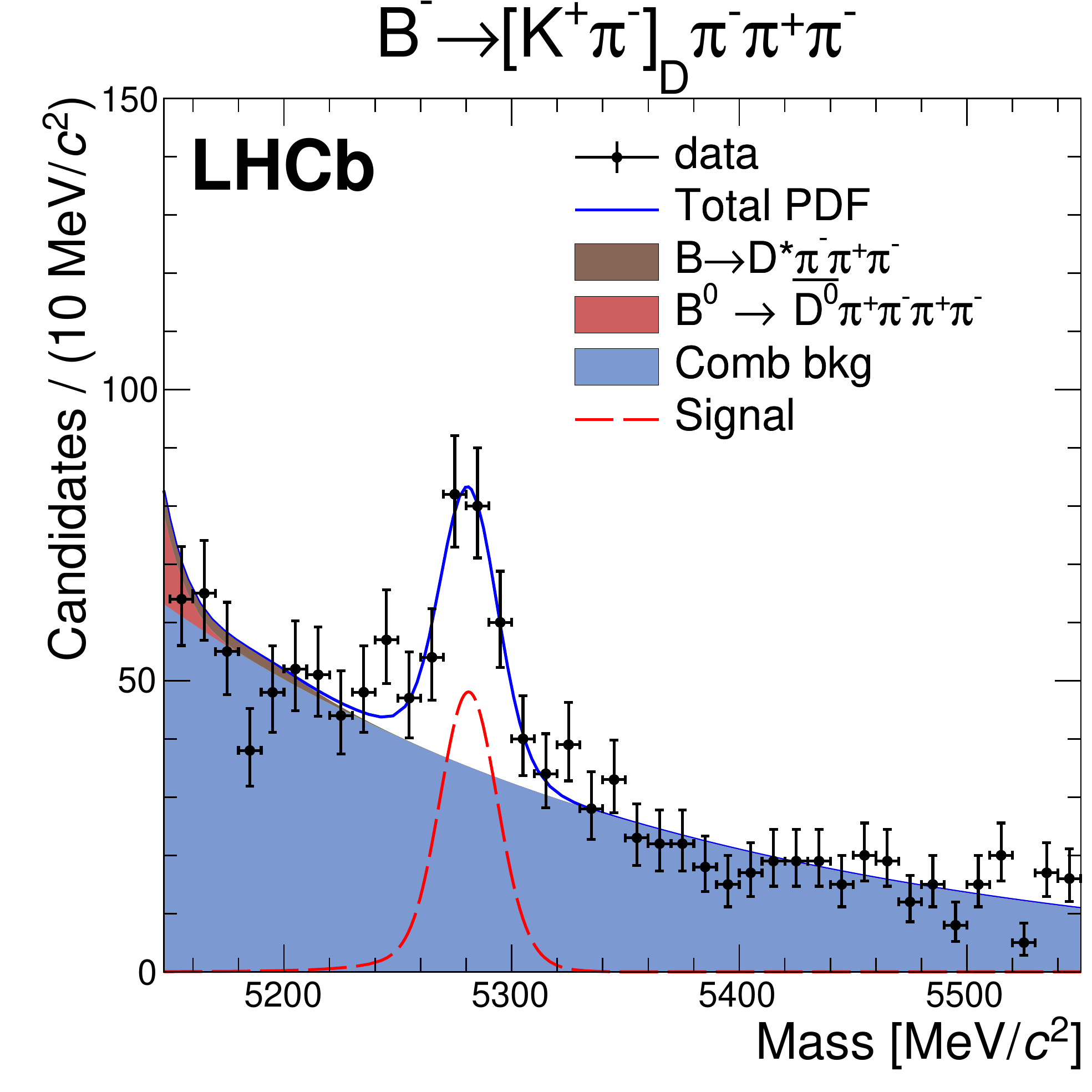}
    \includegraphics[width=0.48\textwidth]{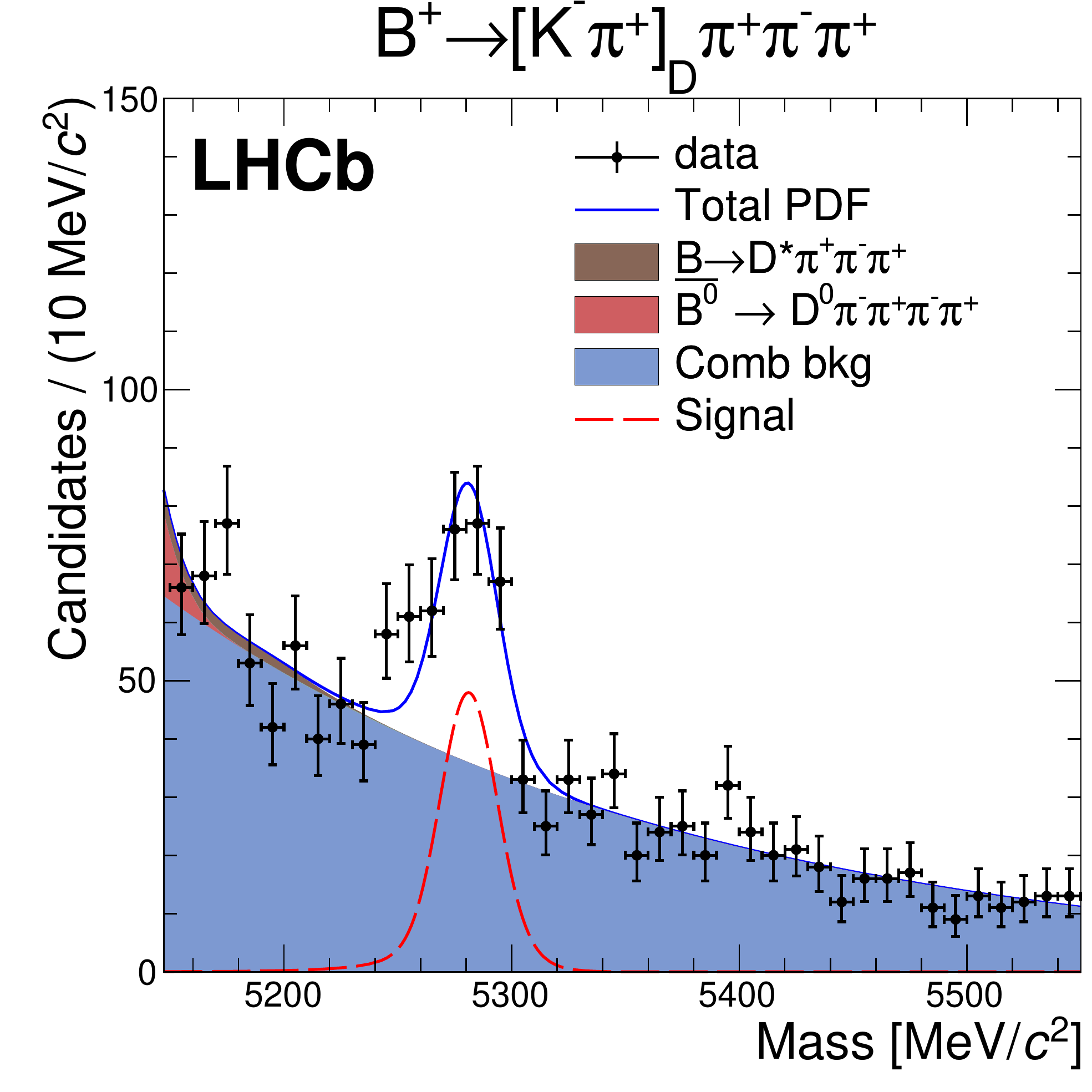}
    \caption{\small{Mass distributions of $\Bm\to DX_d^-$ candidates using the ADS selections, for (top left) $\Bm\to [\Km\pip]_DX_d^-$,
(top right) $\Bp\to [\Kp\pim]_DX_d^+$, (bottom left) $\Bm\to [\Kp\pim]_DX_d^-$, and (bottom right) $\Bp\to [\Km\pip]_DX_d^+$.}}
\label{fig:adsfits_cf}
\end{figure}

\begin{figure}
    \centering
    \includegraphics[width=0.42\textwidth]{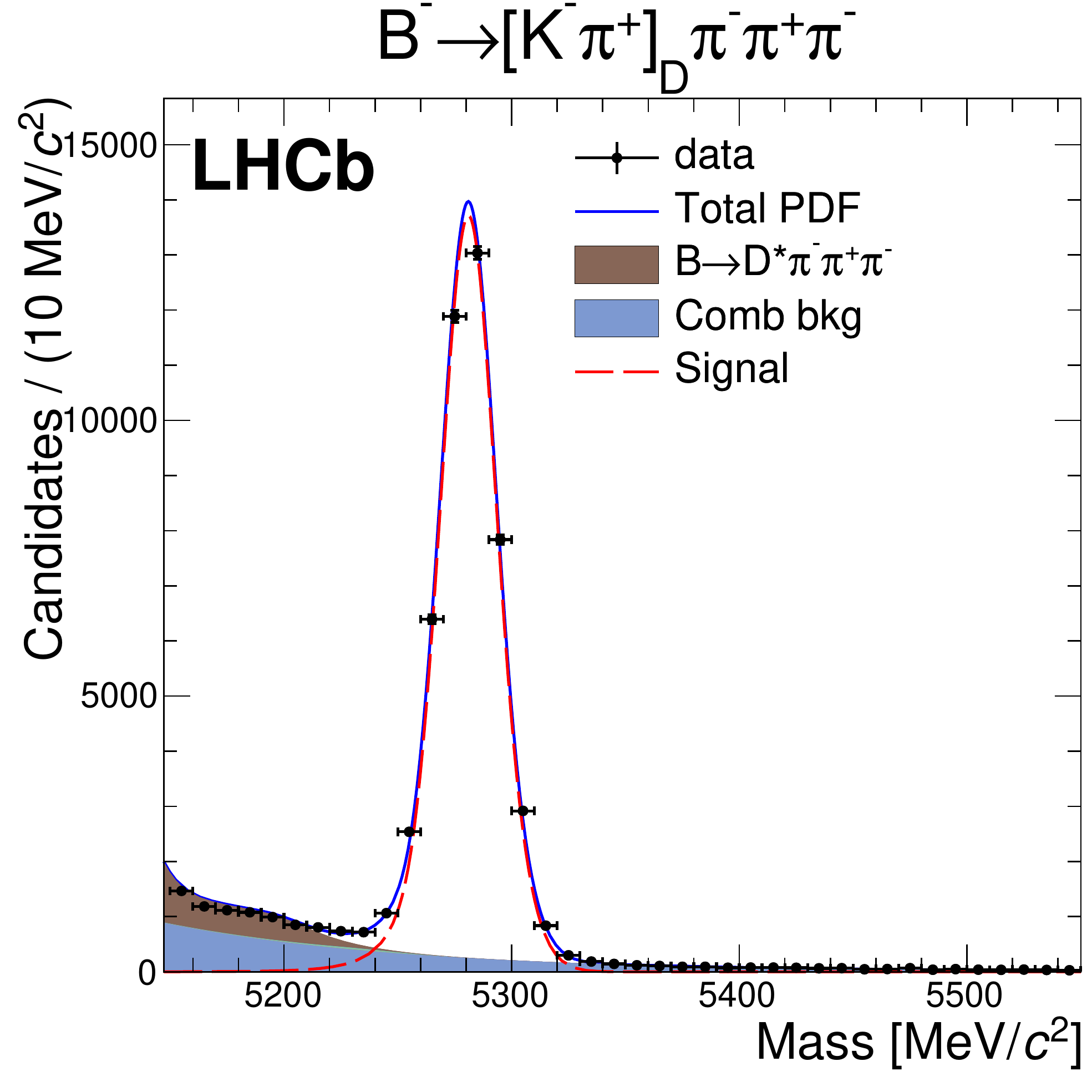}
    \includegraphics[width=0.42\textwidth]{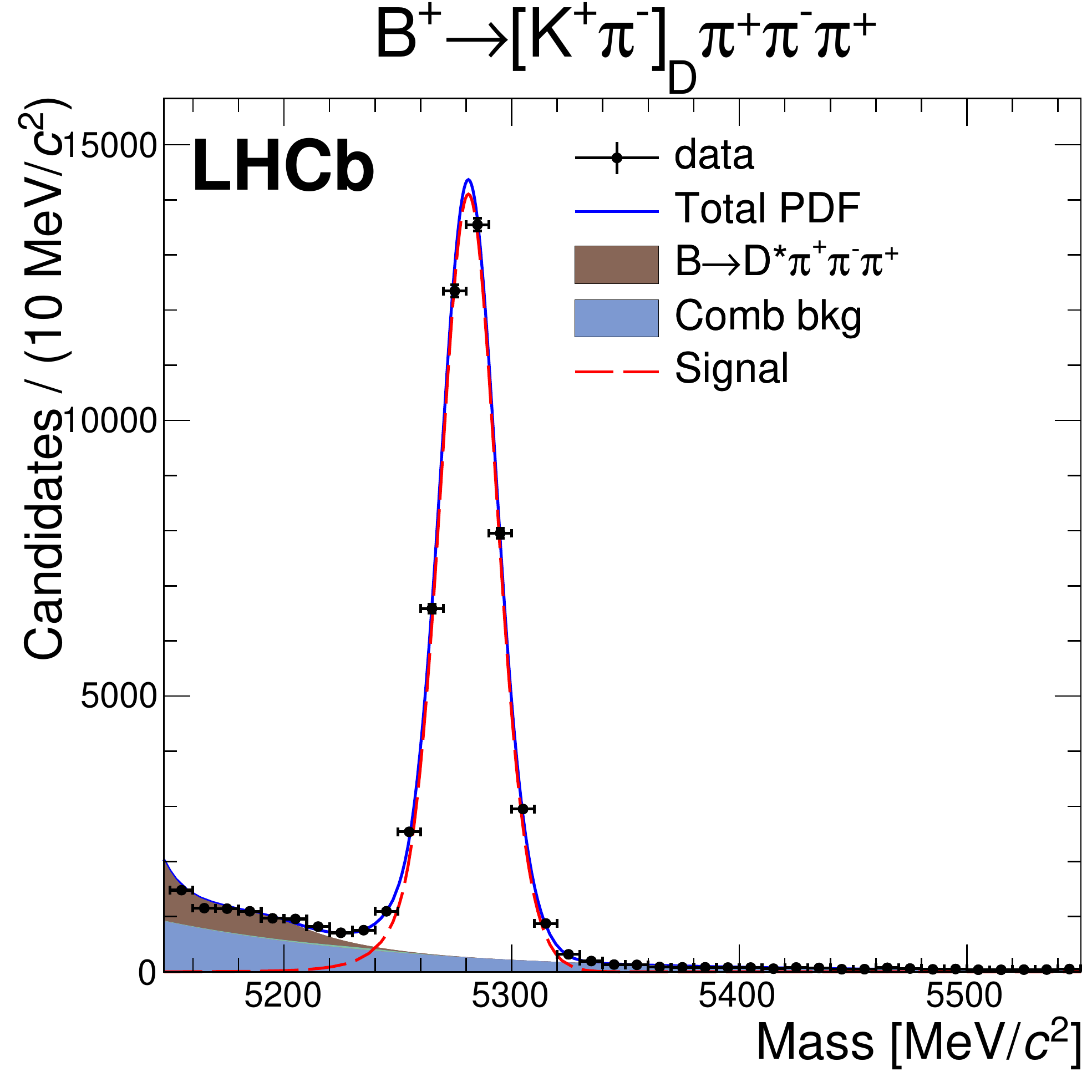}
    \includegraphics[width=0.42\textwidth]{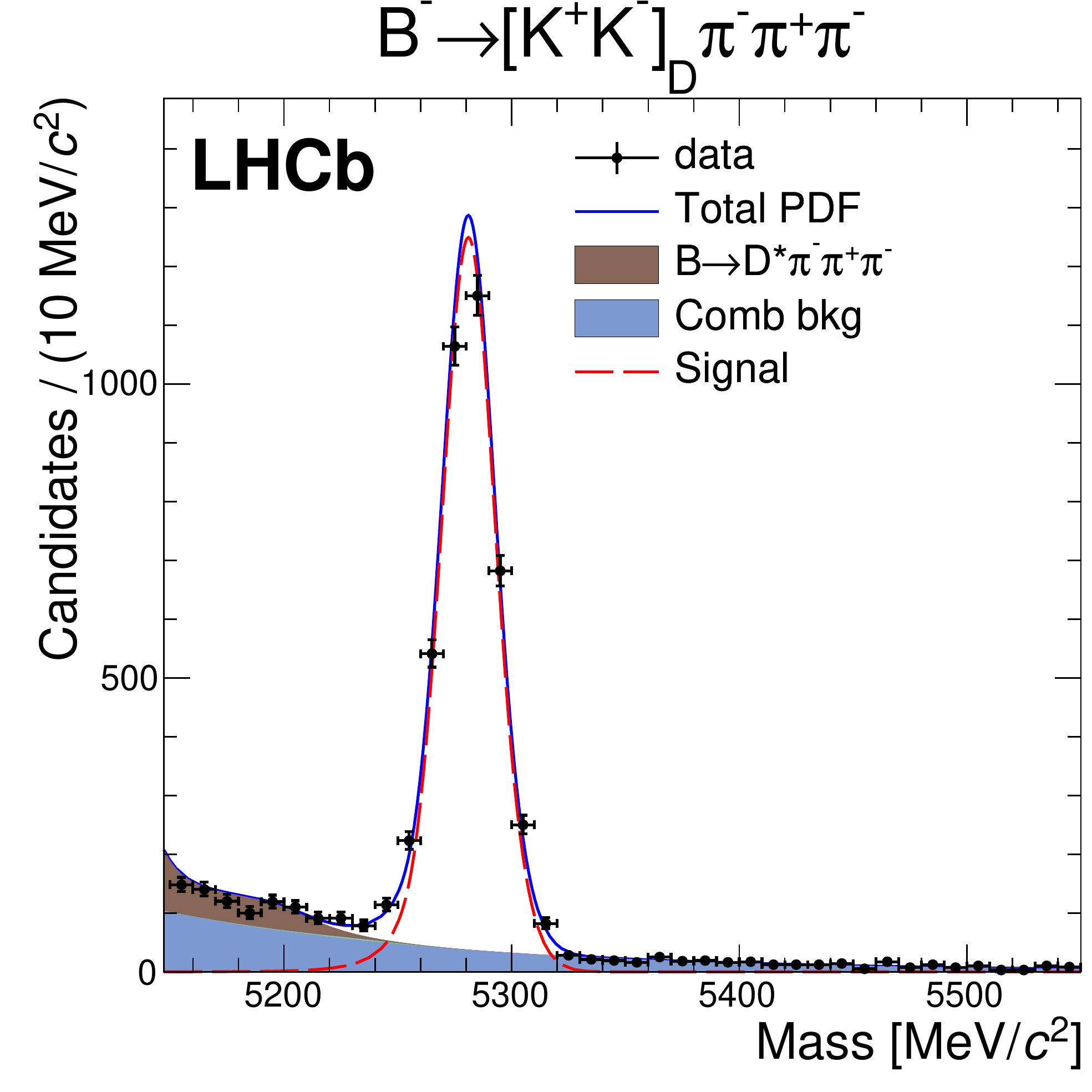}
    \includegraphics[width=0.42\textwidth]{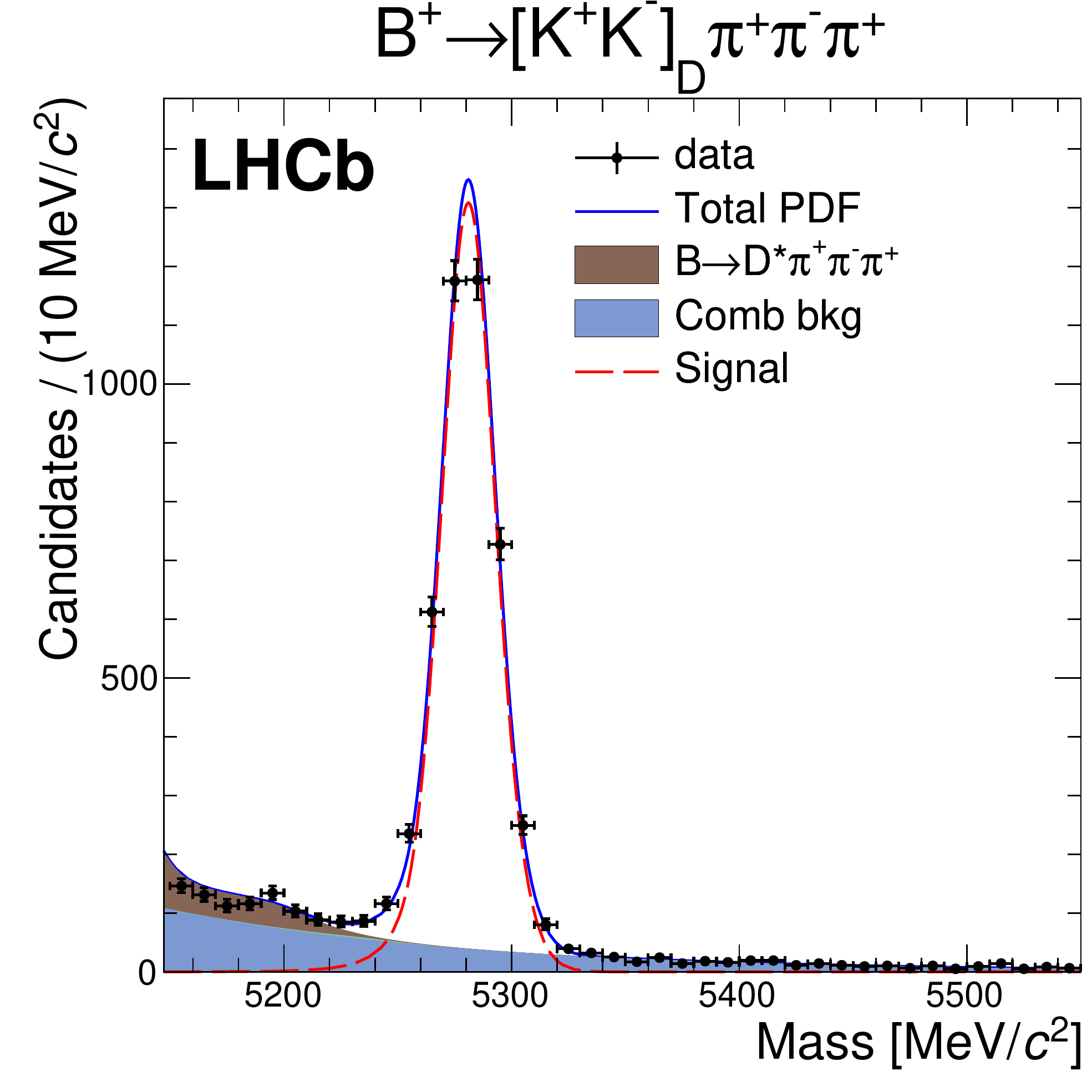}
    \includegraphics[width=0.42\textwidth]{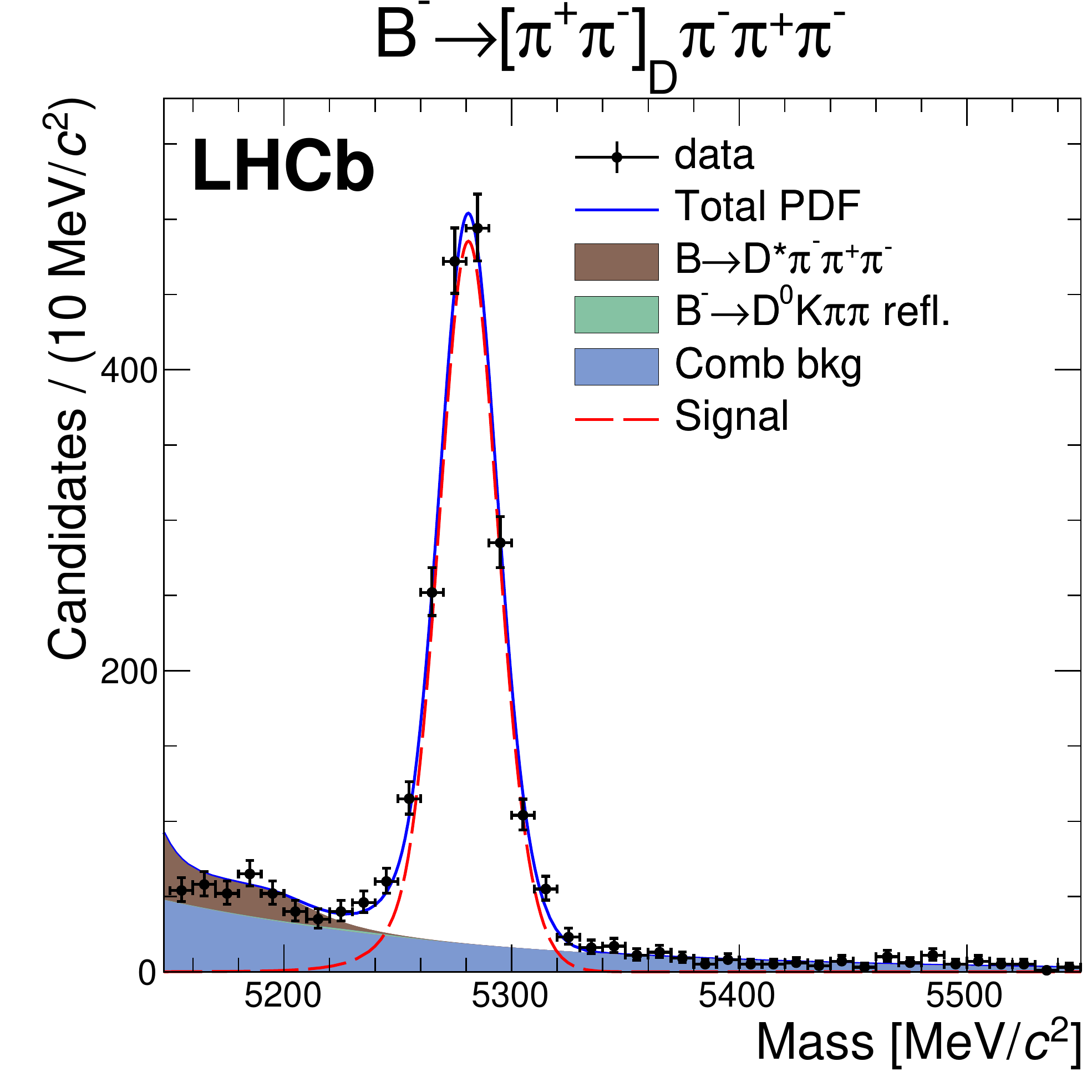}
    \includegraphics[width=0.42\textwidth]{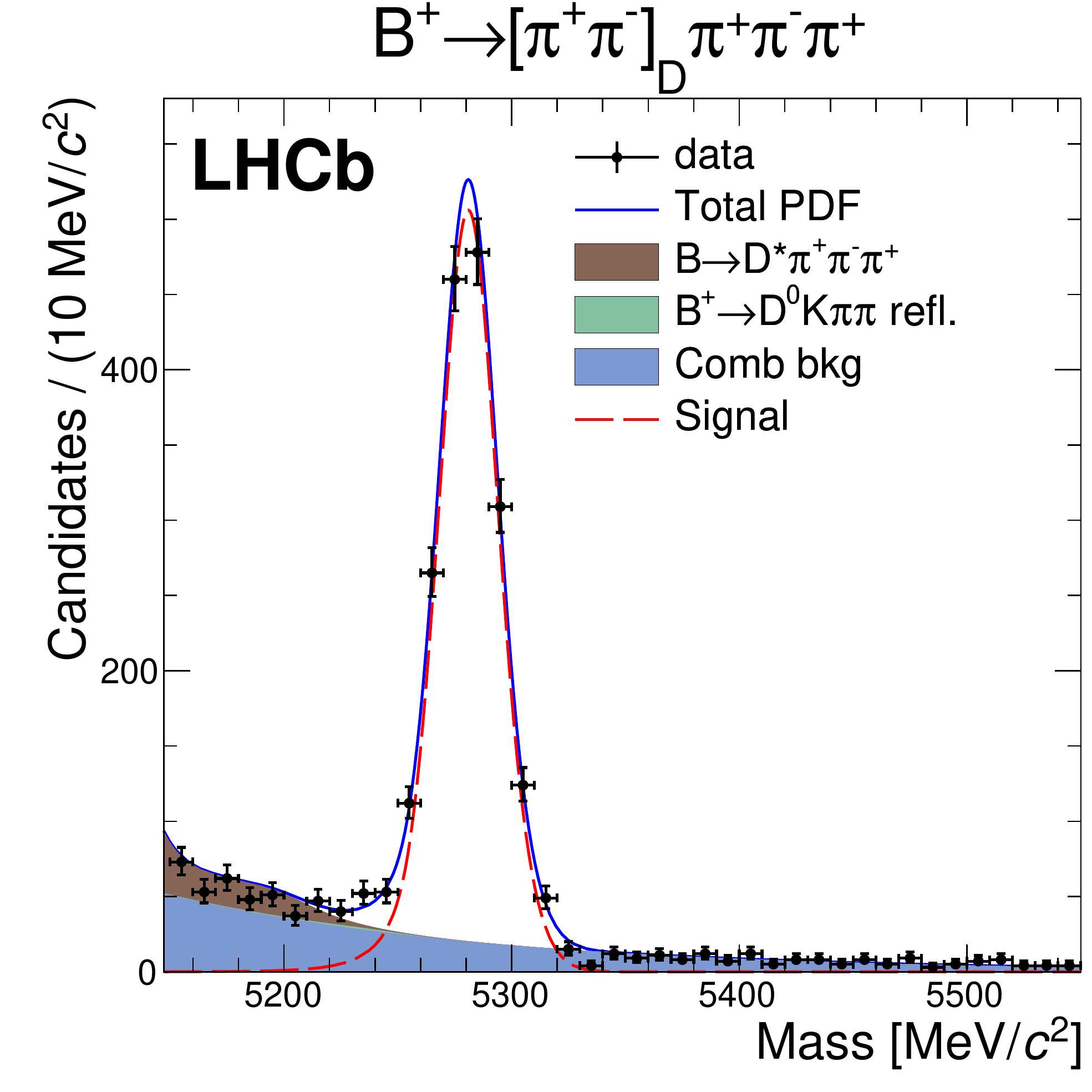}
    \caption{\small{Mass distributions of $\Bm\to DX_d^-$ candidates using the GLW selections, for (top left) $\Bm\to [\Km\pip]_DX_d^-$,
      (top right) $\Bp\to [\Kp\pim]_DX_d^+$, (middle left) $\Bm\to [\Kp\Km]_DX_d^-$, and (middle right) $\Bp\to [\Kp\Km]_DX_d^+$, 
        (bottom left) $\Bm\to [\pip\pim]_DX_d^-$, and 
      (bottom right) $\Bp\to[\pip\pim]_DX_d^+$}}
\label{fig:glwfits_cf}
\end{figure}

The invariant mass spectra for the  $\Bm\to DX_s^-$ ADS and GLW signal modes are shown in Figs.~\ref{fig:adsfits_cs} and~\ref{fig:glwfits_cs}, 
with the corresponding spectra for the $\Bm\to DX_d^-$ normalization modes in Figs.~\ref{fig:adsfits_cf} and~\ref{fig:glwfits_cf}.
Results from the fits are superimposed along with the various signal and background components. 
The fitted yields in the ADS and GLW modes
are given in Tables~\ref{tab:adsyields} and~\ref{tab:glwyields}.

\begin{table*}[tb]
\begin{center}
\caption{\small{Fitted yields in the ADS  modes with $f=K\pi$, for the signal and corresponding normalization modes.}}
\begin{tabular}{lcc}
\hline\hline
Decay mode                             &  $\Bm$ yield                       & $\Bp$ yield \\
                                       &   ($N^f_{{\rm fit}, X_d^{-}}$)    & ($N^f_{{\rm fit}, X_d^{+}}$) \\   
\\[-2.5ex]
\hline
\\[-2.5ex]
$B^{\pm}\to DX_d^{\pm}$, $D\to\Km\pip$ &  $36\,956\pm214$ &  $37\,843\pm219$ \\
\\[-2.5ex]
$B^{\pm}\to DX_d^{\pm}$, $D\to\Kp\pim$ &  $~~161\pm20$ &  $~~162\pm20$ \\
\hline
\\[-2.5ex]
                                      &   ($N^f_{{\rm fit}, X_s^{-}}$)    & ($N^f_{{\rm fit}, X_s^{+}}$)\\
\\[-2.5ex]
\hline
\\[-2.5ex]
$B^{\pm}\to DX_s^{\pm}$, $D\to\Km\pip$ &  $1234\pm37$ &  $1226\pm37$ \\
$B^{\pm}\to DX_s^{\pm}$, $D\to\Kp\pim$ &  $~\,13.0\pm5.3$ &  $~~~6.6\pm4.0$ \\
\hline\hline
\end{tabular}
\label{tab:adsyields}
\end{center}
\end{table*}

\begin{table*}[tb]
\begin{center}
\caption{\small{Fitted yields used in the GLW analysis with $f=K^{\pm}\pi^{\pm},~\Kp\Km$ and $\pip\pim$, for the
signal and corresponding normalization modes.}}
\begin{tabular}{lcc}
\hline\hline
Decay mode                             &   $\Bm$ yield    & $\Bp$ yield \\
                                        &    ($N^f_{{\rm fit}, X_d^{-}}$)    & ($N^f_{{\rm fit}, X_d^{+}}$) \\
\\[-2.5ex]
\hline
\\[-2.5ex]
$B^{\pm}\to DX_d^{\pm}$, $D\to\Km\pip$ &  $45\,213\pm226$ &  $46\,488\pm230$ \\
$B^{\pm}\to DX_d^{\pm}$, $D\to\Kp\Km$ &  $3899\pm63$ &  $4084\pm65$\\
$B^{\pm}\to DX_d^{\pm}$, $D\to\pip\pim$ &  $1669\pm38$ &  $1739\pm40$\\
\hline
\\[-2.5ex]
                                        &    ($N^f_{{\rm fit}, X_s^{-}}$)    & ($N^f_{{\rm fit}, X_s^{+}}$) \\
\\[-2.5ex]
\hline
\\[-2.5ex]
$B^{\pm}\to DX_s^{\pm}$, $D\to\Km\pip$ &  $1699\pm47$ &  $1744\pm47$  \\
$B^{\pm}\to DX_s^{\pm}$, $D\to\Kp\Km$ &  $\,~155\pm14$ &  $\,~171\pm14$  \\
$B^{\pm}\to DX_s^{\pm}$, $D\to\pip\pim$ &  $\,~59\pm9$ &  $\,~70\pm9$  \\
\hline\hline
\end{tabular}
\label{tab:glwyields}
\end{center}
\end{table*}

\begin{figure}[tb]
    \centering
    \includegraphics[width=0.60\textwidth]{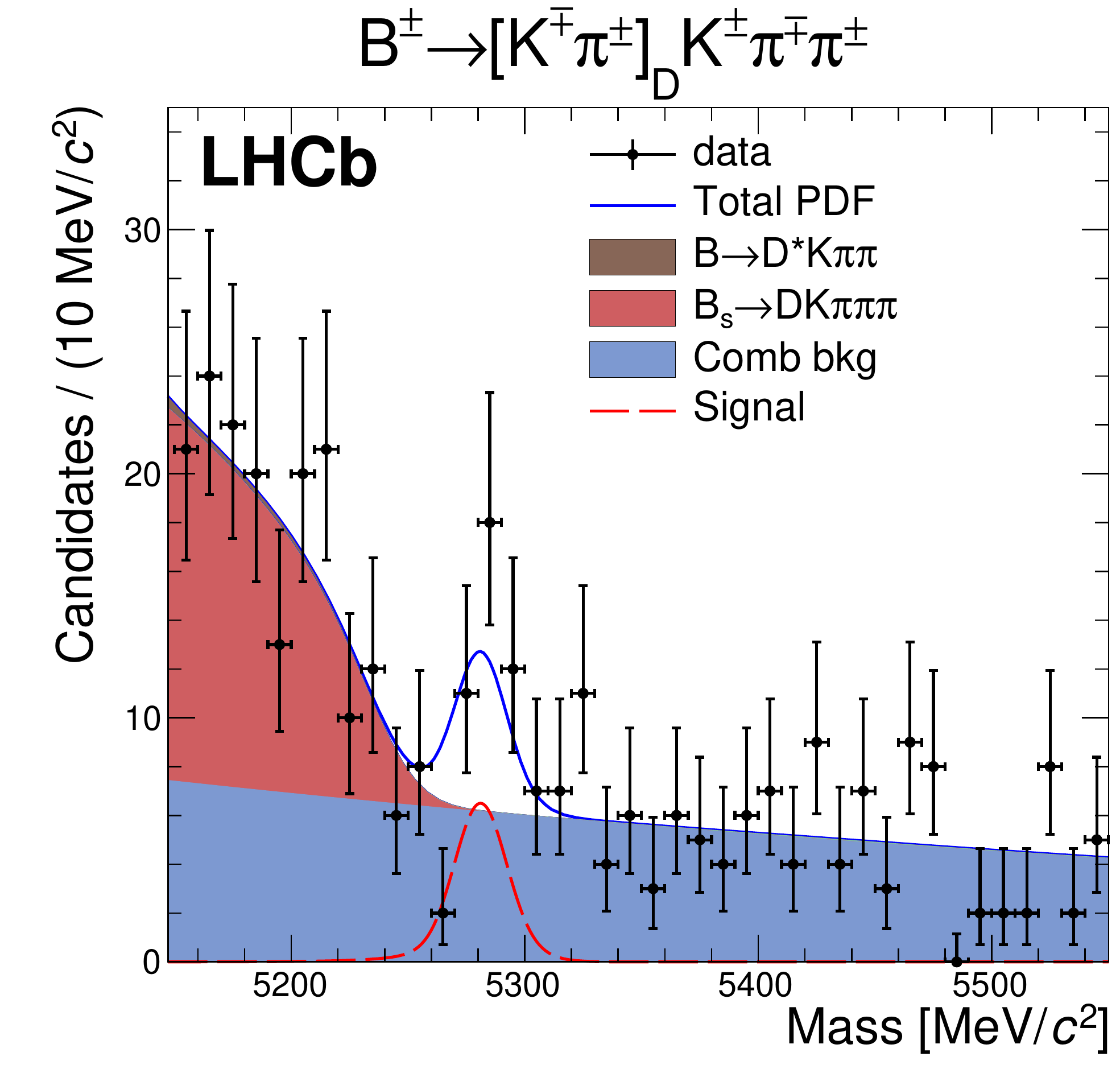}
    \caption{\small{Mass distributions for the suppressed ADS mode, $B^{\pm}\to [K^{\mp}\pi^{\pm}]_DK^{\pm}\pi^{\mp}\pi^{\pm}$ (sum of $\Bp$ and $\Bm$).}}
\label{fig:chgcombined}
\end{figure}

Highly significant signals are seen in all modes, except for the ADS DCS $\Bm\to DX_s^-$ decay. This is the first time these decays have been observed in modes other than the CF $\Dz\to\Km\pip$ decay. 
Figure~\ref{fig:chgcombined} shows the suppressed ADS mode, $B^{\pm}\to D[\Kp\pim]_DK^{\pm}\pi^{\mp}\pi^{\pm}$, summed over
both $B$-meson charge states. The significance of the peak, which exceeds three standard deviations, is discussed later.

\section{Determination of \CP observables}

The \CP observables are obtained by expressing the fitted signal yields in terms of corrected yields and the
\CP parameters. For the decay $B^{\pm}\to f_DX_d^{\pm}$, where $f_D$ is either the ADS CF decay or a \CP eigenstate,
the fitted yields can be written as
\begin{align}
N^f_{{\rm fit}, X_d^{\pm}}  = \frac{1}{2} \Bigg(\frac{N_{{\rm corr}, X_d}^f}{1+F^f_{\Dveto, X_d}} \Bigg)(1\mp\Asy_{{\rm raw},X_d}^f)+C^f_{\nonc,X_d^{\pm}},
\label{eq:cfyield}
\end{align}
\noindent where $N_{{\rm corr}, X_d}^f$ is the total corrected yield (sum of $\Bm$ and $\Bp$), 
$F^f_{\Dveto, X_d}$ are the estimated fractions of signal events removed by the $\Dz$ and $D_{(s)}^{(*)+}$ vetoes, 
$C^f_{\nonc,X_d^{\pm}}$ are the estimated charmless background yields, and $\Asy_{{\rm raw},X_d}^f$ is the raw \CP asymmetry.

The fitted yields in the corresponding $B^{\pm}\to DX_s^{\pm}$ decays are written in terms of the corrected 
$B^{\pm}\to DX_d^{\pm}$ yields in Eq.~\ref{eq:cfyield} and the \CP observable $R^f_{s/d}$ defined in Eqs.~\ref{eq:rglwapp1} and~\ref{eq:rglwapp2}, as
\begin{align}
N^f_{{\rm fit}, X_s^{\pm}} = \frac{1}{2}R^f_{s/d} \Bigg(\frac{N^f_{{\rm corr}, X_d}}{1+F^f_{\Dveto, X_s}}\Bigg)(1\mp\Asy_{{\rm raw},X_s}^f)+C^f_{\nonc,X_s^{\pm}},
\label{eq:csyield}
\end{align}
\noindent where the meaning of the symbols parallels those in Eq.~\ref{eq:cfyield}. 

For the ADS suppressed modes, the four DCS yields $N^{K^{\mp}\pi^{\pm}}_{{\rm fit}, X^{\pm}}$
are expressed in terms of the corrected CF yields,  $N_{{\rm corr}, X_d}^{K^{\pm}\pi^{\mp}}$, as 
\begin{align}
N^{K^{\mp}\pi^{\pm}}_{{\rm fit}, X_d^{\pm}} = \left(R^{X_d^{\pm}}_{{\rm raw}}\right) \Bigg(\frac{N_{{\rm corr}, X_d}^{K^{\pm}\pi^{\mp}}}{1+F^{K^{\mp}\pi^{\pm}}_{\Dveto, X_d}} \Bigg)+C^{K^{\mp}\pi^{\pm}}_{\nonc,X_d^{\pm}}, \\
N^{K^{\mp}\pi^{\pm}}_{{\rm fit}, X_s^{\pm}} = \left(R^{X_s^{\pm}}_{{\rm raw}}\right) \Bigg(\frac{N_{{\rm corr}, X_s}^{K^{\pm}\pi^{\mp}}}{1+F^{K^{\mp}\pi^{\pm}}_{\Dveto, X_s}} \Bigg)+C^{K^{\mp}\pi^{\pm}}_{\nonc,X_s^{\pm}}, 
\label{eq:adsyield}
\end{align}
\noindent where $N_{{\rm corr}, X_s^{\pm}}^{K^{\pm}\pi^{\mp}}=N_{{\rm corr}, X_s}^{K^{\pm}\pi^{\mp}}(1\mp\Asy_{{\rm raw},X_s}^{K^{\pm}\pi^{\mp}})$
gives the corrected yield for the favored $B^{\pm}\to [K^{\pm}\pi^{\mp}]_DX_s^{\pm}$ decays.

The corrections for the $\Dz$ and $D_{(s)}^{(*)+}$ vetoes,  $F^f_{\Dveto, X_{d,s}}$, are determined by interpolating from the mass regions
just above and below the veto region, and lead to corrections that range from 0.6\% to 5.8\% of the expected yield.
Uncertainties on these corrections are considered as sources of systematic uncertainty. Potential contamination
from charmless five-body decays is determined by fitting for a $B^{\pm}$ signal component when the $D$ candidates are taken from the
$\Dz$ mass sideband region, as described previously. 
The charmless contributions are negligible, and the uncertainties are included in the systematic error.
The yields, as determined from the fitted values of the \CP parameters
in Eqs.~\ref{eq:cfyield}-\ref{eq:adsyield}, are given in Tables~\ref{tab:adsyields} and ~\ref{tab:glwyields}.

The raw observables, $\Asy_{{\rm raw},X}^f$ and $R^{X^{\pm}}_{{\rm raw}}$ include small biases due to the 
production asymmetry of $B^{\pm}$ mesons, ${\cal{A}}_{B^{\pm}}$ (affecting $\Asy_{{\rm raw},X}^f$ only), and from 
the detection asymmetries of kaons and pions, ${\cal{A}}_{K}$ and ${\cal{A}}_{\pi}$. The corrected quantities 
are then computed according to

\begin{align}
\Asy^{\Kp\Km}_{X_d} &=  \Asy^{\Kp\Km}_{{\rm raw},X_d} - {\cal{A}}_{B^{\pm}} - {\cal{A}}_{\pi}, \\
\Asy^{\pip\pim}_{X_d} &=  \Asy^{\pip\pim}_{{\rm raw},X_d} - {\cal{A}}_{B^{\pm}} - {\cal{A}}_{\pi}, \\
\Asy^{\Km\pip}_{X_d} &=  \Asy^{\Km\pip}_{{\rm raw},X_d} - {\cal{A}}_{B^{\pm}} - {\cal{A}}_{K}, \\
\Asy^{\Kp\Km}_{X_s} &=  \Asy^{\Kp\Km}_{{\rm raw},X_s} - {\cal{A}}_{B^{\pm}} - {\cal{A}}_{K}, \\
\Asy^{\pip\pim}_{X_s} &=  \Asy^{\pip\pim}_{{\rm raw},X_s} - {\cal{A}}_{B^{\pm}} - {\cal{A}}_{K}, \\
\Asy^{\Km\pip}_{X_s} &=  \Asy^{\Km\pip}_{{\rm raw},X_s} - {\cal{A}}_{B^{\pm}} - 2{\cal{A}}_{K} + {\cal{A}}_{\pi}, \\
R^{X_d^+} &= R^{X_d^+}_{{\rm raw}}(1-2{\cal{A}}_{K}+2{\cal{A}}_{\pi}), \\
R^{X_d^-} &= R^{X_d^-}_{{\rm raw}}(1+2{\cal{A}}_{K}-2{\cal{A}}_{\pi}), \\
R^{X_s^+} &= R^{X_s^+}_{{\rm raw}}(1-2{\cal{A}}_{K}+2{\cal{A}}_{\pi}), \\
R^{X_s^-} &= R^{X_s^-}_{{\rm raw}}(1+2{\cal{A}}_{K}-2{\cal{A}}_{\pi}). 
\end{align}
\noindent The pion detection asymmetry of ${\cal{A}}_{\pi}=0.000\pm0.003$ is
obtained by reweighting the measured $\pi^{\pm}$ detection efficiencies~\cite{LHCb-PAPER-2012-009}
with the expected momentum spectrum for signal pions. The kaon detection efficiency of 
${\cal{A}}_{K}=-0.011\pm0.004$ is obtained by reweighting the measured $\Km\pi$ detection asymmetry~\cite{LHCb-PAPER-2014-013}
using the momentum spectrum of signal kaons, and then subtracting the above pion detection asymmetry.
For the production asymmetry, the value ${\cal{A}}_{B^{\pm}}=-0.008\pm0.007$ is used~\cite{LHCb-PAPER-2012-055}, based on the
measured raw asymmetry in $B^{\pm}\to\jpsi K^{\pm}$ decays~\cite{LHCb-PAPER-2011-024} and on simulation.

\subsection{Systematic uncertainties}
Most potential systematic uncertainties on the observables are expected to cancel in either the asymmetries or ratios that are measured. The
systematic uncertainties that do not cancel completely are summarized in Table~\ref{tab:syst}.
The PID and trigger asymmetries are evaluated using measured kaon and pion efficiencies from $\Dstarp\to\Dz\pip$ calibration samples in data 
that are identified using only the kinematics of the decay. The efficiencies for the $\Bp$ and $\Bm$ signal decays are then obtained by 
reweighting the kaon and pion efficiencies using simulated $B^{\pm}\to DX^{\pm}$ decays to represent the properties of signal data.
We find no significant charge asymmetry with respect to
the PID requirements, and use ${\cal{A}}_{h}^{\rm PID}=0.000\pm0.006$, where the uncertainty is dominated by the finite sample sizes of the
simulated signal decays in the reweighting. The asymmetry of the hardware trigger is assessed using measured hadron trigger efficiencies
in $\Dstarp\to\Dz\pip,~\Dz\to\Km\pip$ decays,
reweighted to match the momentum spectrum of tracks from signal decays. Defining the $B^{\pm}$ hadron trigger efficiency as $\epsilon_{B^{\pm}}$,
the charge asymmetry of the trigger $(\epsilon_{\Bm}-\epsilon_{\Bp})/(\epsilon_{\Bm}+\epsilon_{\Bp})$
varies from $0.000\pm0.003$ for $\Bm\to[\Kp\Km]_{D}X_s^-$ to $0.007\pm0.003$ for $\Bm\to[\Kp\pip]_{D}X_s^-$. These values are applied 
as corrections.

\begin{table*}[tb]
\begin{center}
\caption{\small{Systematic uncertainties, in percent,  on the fitted parameters.}}
\begin{tabular}{lcccccccc}
\hline\hline
\\[-2.5ex]
Source              & \multicolumn{2}{c}{${\cal{A}}(B^{\pm}\to D X_d^{\pm})$} & \multicolumn{2}{c}{${\cal{A}}(B^{\pm}\to D X_s^{\pm})$} & \multicolumn{2}{c}{$R_{\CP+}$} & $R_{d}^{\pm}$ & $R_{s}^{\pm}$ \\
~~~~~~~~~~~~~$D\to$   & $h^+h^-$ & $K\pi$ & $h^+h^-$ & $K\pi$ & $\Kp\Km$ & $\pip\pim$ & $K\pi$ & $K\pi$ \\
\hline
${\cal{A}}_{B^{\pm}}$ & 0.7 & 0.7 & 0.7 & 0.7 &-- &-- &-- &--  \\
${\cal{A}}_{K}$     & -- &  0.4 & 0.4 & 0.8 &-- &-- & 0.7 & 0.7 \\
${\cal{A}}_{\pi}$   & 0.3 & -- & -- & 0.3 & -- & -- & 0.6 & 0.3 \\
Trigger            & 0.4 & 0.4 & 0.4 & 0.4 & 1.5 & 1.5 & 1.5 & 1.5 \\
PID                & 0.6 & 0.6 & 0.6 & 0.6 & 1.2 & 1.2 & 1.2 & 1.2 \\
Signal model       &-- &-- &-- &-- & 1.1 & 1.1 &-- &-- \\
Bkgd. model       &-- &-- &-- &-- & 1.6 & 1.6 & 4.0 & 10.0  \\
Charmless back.   &-- &-- &-- &-- & 1.0 & 1.0 & 1.0 & 1.0 \\
Cross-feed        & --&-- &-- &-- & 1.0 & 1.0 & 1.0 & 1.0\\
$D$ vetoes        &-- & --&-- &-- & 1.0 & 1.7 & 1.0 & 1.0\\
$R_{\CP+}$ approx.      &-- & --&-- &-- & 1.0 & 1.0 &-- &-- \\
\hline
Total          & 1.0 & 1.1 & 1.1 & 1.3 & 3.4 & 3.8 & 4.9 & 10.4 \\     
\hline\hline
\end{tabular}
\label{tab:syst}
\end{center}
\end{table*}

On $R_{\CP+}$ and $R^{X^{\pm}}$, we have either a double-ratio or a ratio of final states with identical particles (apart from
the charges), and therefore there is a high degree of cancellation of potential systematic uncertainties. We expect that
for these ratios, the relative trigger efficiencies would yield a value close to unity. After reweighting the measured trigger efficiencies
according to the kinematical properties of signal decays (obtained from simulation), we find that the ratios of trigger efficiencies are 
within 1.5\% of unity,
which is assigned as a systematic uncertainty. Using an analogous weighting procedure to the measured PID efficiencies, we find that the
relative PID efficiency is equal to unity to within 1.2\%, which is assigned as a systematic uncertainty.

We also consider uncertainty from the signal model, the background model, 
the charmless contamination, the $D$ vetoes, and the detection asymmetries. 
For the signal model uncertainty, all of the fixed signal shape parameters are varied
by one standard deviation, and the resulting changes in the \CP parameters are added in quadrature to
obtain the total signal shape uncertainty (1.1\%). For the background-related uncertainties, we consider a polynomial
function for the combinatorial background, and vary the fixed background shape parameters of the specific $b$-hadron 
backgrounds within their uncertainties, and add the deviations from the nominal result in quadrature (1.6\%). 
For the ADS-suppressed modes, larger uncertainties are assigned based on an incomplete understanding of the 
contributions to the low mass $\bar{B}^{0}_{(s)}\to\Dz X$ background.

The charmless backgrounds are all consistent with zero, and the uncertainty is taken from fits to the $D$ sideband regions (1.0\%).
Uncertainties due to the cross-feed contributions (such as $\Bm\to DX_d^-$ reconstructed as $\Bm\to DX_s^-$) are assessed 
using simulated experiments, by simulating the mass distributions with a larger cross-feed and fitting with the nominal value 
(1.0\%). The uncertainties due to vetoing potential contributions from other $D$ mesons are assessed
by interpolating the mass spectrum just above and below the veto region into the veto region.
The associated uncertainties are all at the 1.0\% level, except for the 
the $B\to [\pip\pim]_DX_s^-$ mode, which has an uncertainty of 1.7\%.

The uncertainties on the ratios $R_{s/d}^{h^+h^{\prime-}}$ and $R^{X_{s,d}^-}$ are each summed in quadrature, 
giving total uncertainties in the range of $(3.4-10.4)\%$, depending on the mode. 

\section{Results and summary}

The resulting values for the \CP observables are
\begin{align*}
R^{\Kp\Km}_{\CP+} &= 1.043\pm0.069\pm0.034,\\
R^{\pip\pim}_{\CP+} &= 1.035\pm0.108\pm0.038,\\
\Asy^{\Kp\Km}_{X_d} &= -0.019\pm0.011\pm0.010, \\
\Asy^{\pip\pim}_{X_d} &= -0.013\pm0.016\pm0.010, \\
\Asy^{\Km\pip}_{X_d} &=  -0.002\pm0.003\pm0.011, \\
R^{X_d^+} &= (43.2\pm5.3\pm2.1)\times10^{-4}, \\
R^{X_d^-} &= (42.1\pm5.3\pm2.1)\times10^{-4}, \\
\Asy^{\Kp\Km}_{X_s} &= -0.045\pm0.064\pm0.011, \\
\Asy^{\pip\pim}_{X_s} &= -0.054\pm0.101\pm0.011, \\
\Asy^{\Km\pip}_{X_s} &=  0.013\pm0.019\pm0.013, \\
R^{X_s^+} &= (107_{-44}^{+60}\pm11)\times10^{-4}~~~[~ < 0.018~\mbox{\rm at 95\% CL}~], \\
R^{X_s^-} &= (53_{-42}^{+45}\pm6)\times10^{-4}~~~[~ < 0.012~\mbox{\rm at 95\% CL}~].
\end{align*}
\noindent The values of $R_{\CP+}$ are averaged to obtain
\begin{align*}
R_{\CP+} &= 1.040\pm0.064, 
\end{align*}
\noindent where the uncertainty includes both statistical and systematic sources, as well as the correlations
between the latter.

The significances of the suppressed ADS modes are determined by computing the ratio of log-likelihoods, 
$\sqrt{2\log({\cal{L}}_0/{\cal{L}}_{\rm min})}$, after convolving ${\cal{L}}$ with the systematic uncertainty.
From the value of ${\cal{L}}$ at the minimum (${\cal{L}}_{\rm min}$), and the value at $R^{X_s^{\pm}}=0$ (${\cal{L}}_{0}$),
the significances of the non-zero values for $R^{X_s^-}$ and $R^{X_s^+}$ are found to be 2.0$\sigma$ and 3.2$\sigma$, respectively. 
The overall significance of the observation
of the ADS suppressed mode is obtained by adding the log-likelihoods, resulting in a significance of 3.6 standard deviations. 
This constitutes the first evidence of the ADS suppressed mode in $\Bm\to D\Km\pip\pim$ decays.

For completeness, we also compute the related observables $R_{\rm ADS}$ and $\Asy_{\rm ADS}$, which are commonly used. For the
$\Bm\to DX_s^-$ modes, the values are
\begin{align*}
R_{\rm ADS}^{X_s}&\equiv (R^{X_s^-}+R^{X_s^+})/2 = (85^{+36}_{-33})\times10^{-4},\\
\Asy_{\rm ADS}^{X_s}&\equiv \frac{R^{X_s^-}-R^{X_s^+}}{R^{X_s^-}+R^{X_s^+}} =-0.33^{+0.36}_{-0.34}.
\end{align*}
\noindent For the favored modes, the corresponding values are
\begin{align*}
R_{\rm ADS}^{X_d}&\equiv (R^{X_s^-}+R^{X_s^+})/2 = (42.7\pm5.6)\times10^{-4},\\
\Asy_{\rm ADS}^{X_d}&\equiv \frac{R^{X_s^-}-R^{X_s^+}}{R^{X_s^-}+R^{X_s^+}} =-0.013\pm0.087.
\end{align*}
The averages are computed using the asymmetric uncertainty distributions, and
include both statistical and systematical sources.

To assess the constraints on $\gamma$ that
these observables provide, they have been implemented in the fitter for $\gamma$ described in Ref.~\cite{LHCb-PAPER-2013-020,*LHCb-CONF-2014-004}.
Two fits are performed, one that uses only information from $\Bm\to\ DX_s^-$, and a second that uses the observables from both
$\Bm\to\ DX_s^-$ and $\Bm\to\ DX_d^-$ decays.
In both fits, the parameters from the $D$-meson system, $r_D$, $\delta_D^{K\pi}$, $x_D$, $y_D$, $A^{\rm dir}_{CP}(\Kp\Km)$, 
and $A^{\rm dir}_{CP}(\pip\pim)$,
are constrained in an analogous way to what was done for the $\Bm\to D\Km$ and $\Bm\to D\pim$ case~\cite{LHCb-PAPER-2013-020,*LHCb-CONF-2014-004}. 
The four parameters $r_B$, $\delta_B$, $\kappa$ and $\gamma$
are freely varied in each fit. In the combined fit, 
three additional strong parameters, $r_B^{DX_d}$, $\delta_B^{DX_d}$, $\kappa^{DX_d}$ are included, which are analogous to 
those that apply to the $\Bm\to\ DX_s^-$ decay.

\begin{figure}
    \centering
    \includegraphics[width=0.48\textwidth]{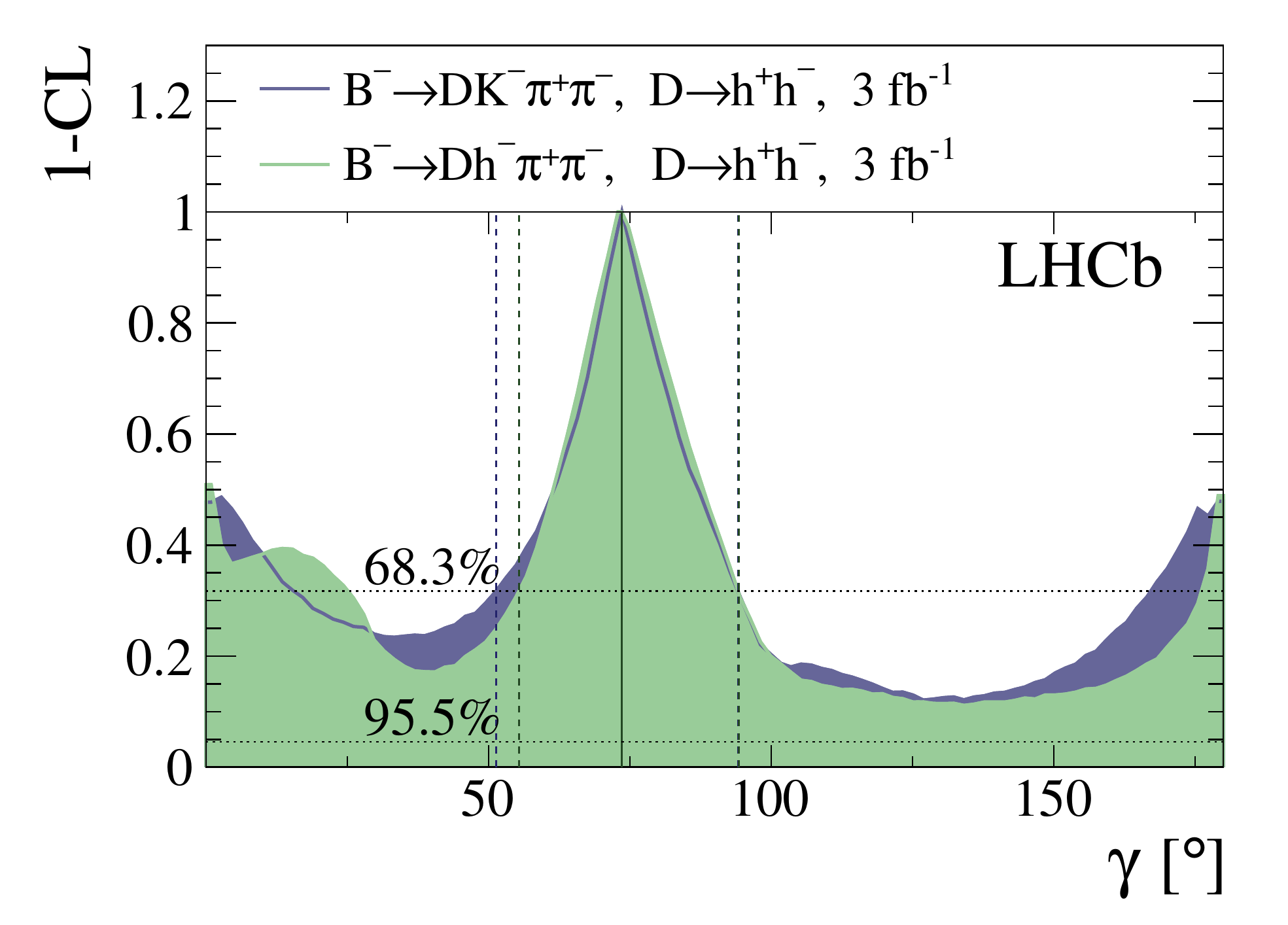}
    \includegraphics[width=0.48\textwidth]{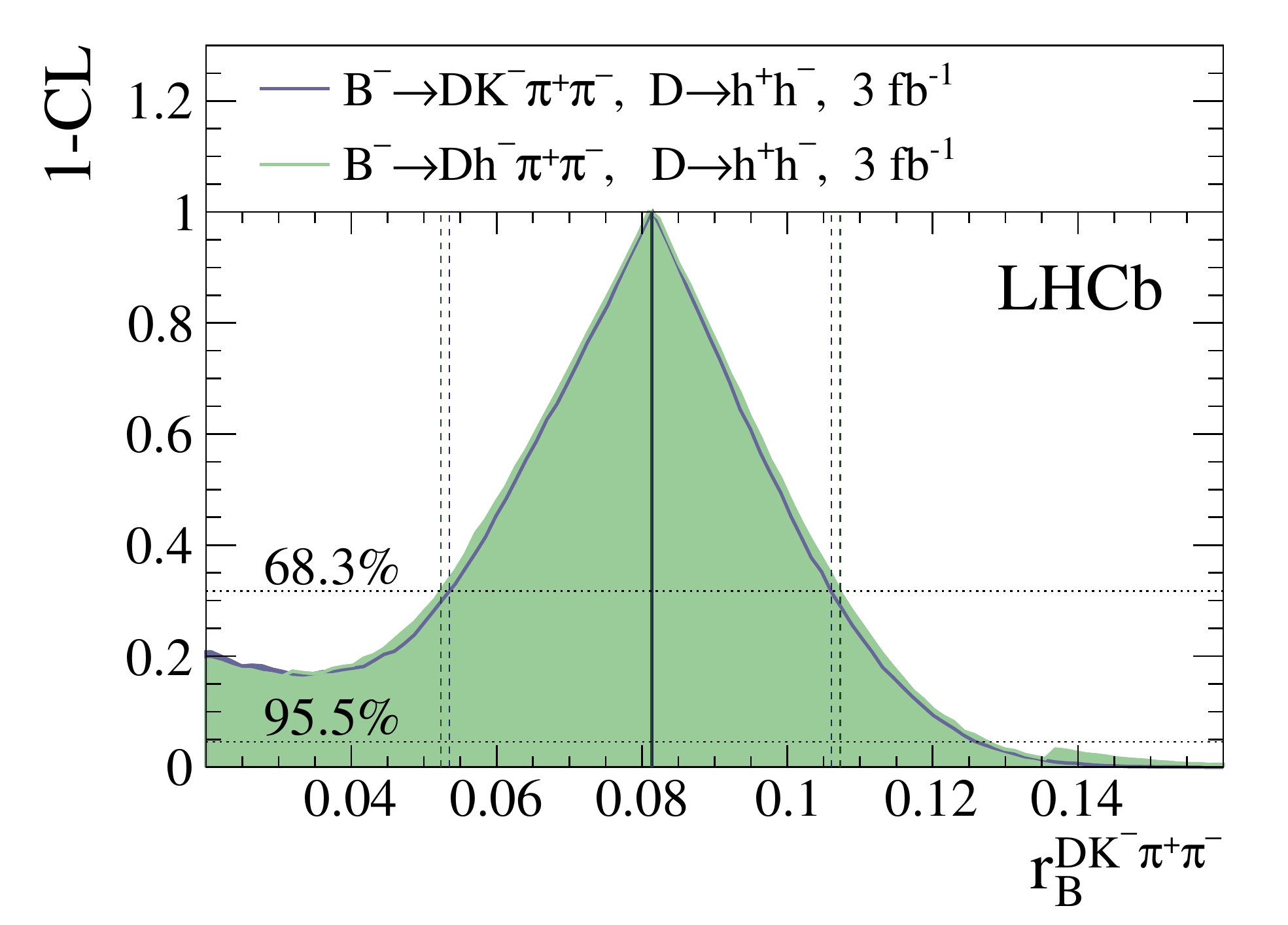}
    \includegraphics[width=0.58\textwidth]{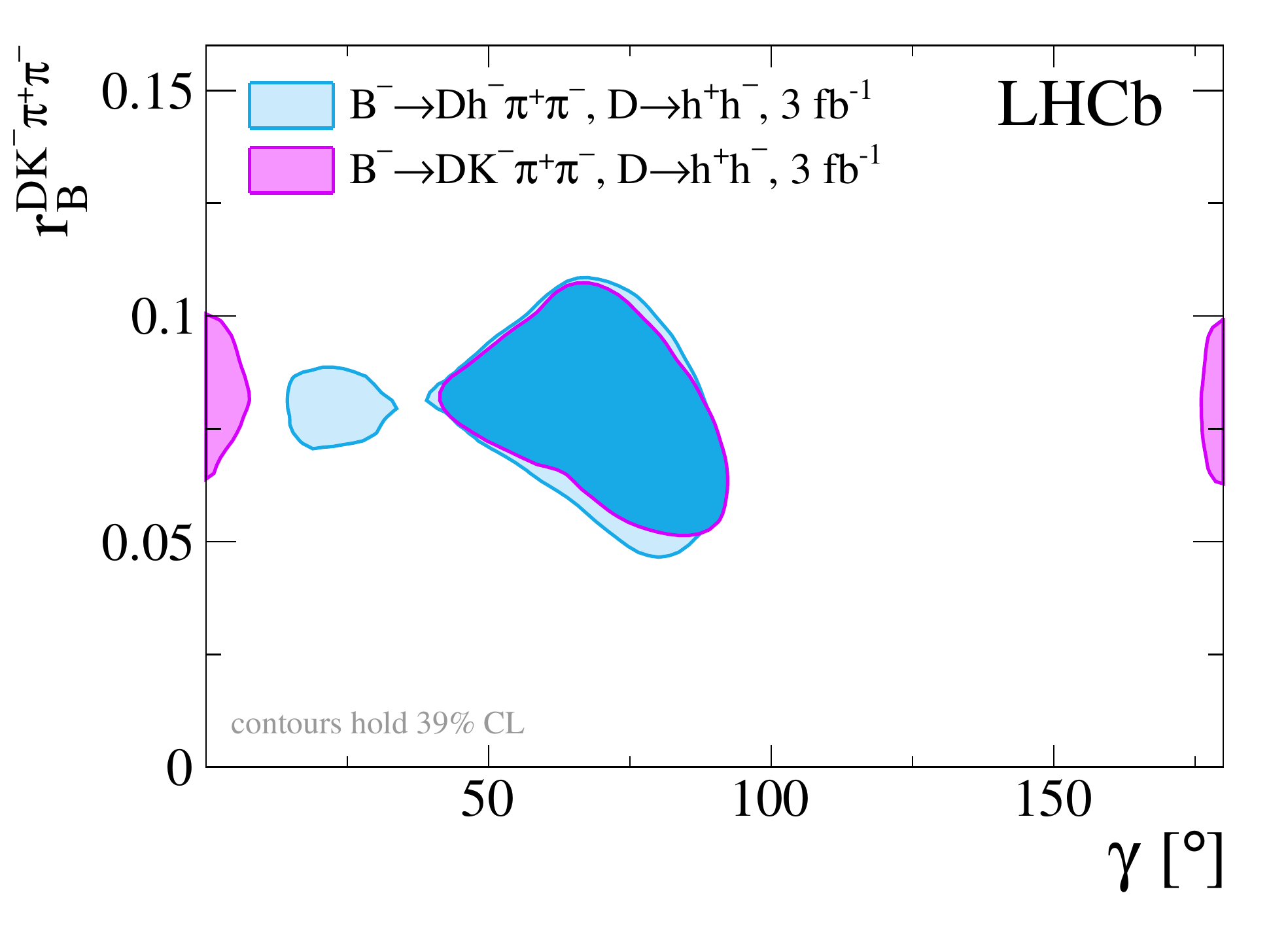}
    \caption{\small{Projections of 1$-$CL versus (left) $\gamma$, (right) $r_B^{\Bm\to D\Km\pip\pim}$, and (bottom) $r_B^{\Bm\to D\Km\pip\pim}$ versus $\gamma$,
using only $\Bm\to\ D\Km\pip\pim$ decays, and the combination of 
$\Bm\to\ D\Km\pip\pim$ and $\Bm\to\ D\pim\pip\pim$ decays. The 68.3\% and 95.5\% confidence level (CL) limits are indicated for
the $\gamma$ and $r_B$ projections. The 39\% level contours in $r_B^{\Bm\to D\Km\pip\pim}$ versus $\gamma$ correspond to the 
68.3\% level contours in the one-dimensional projections.}} 
\label{fig:gammaCombination_D3h}
\end{figure}

The projections of the fit results for $\gamma$, $r_B$ and $r_B$ versus $\gamma$, are shown in Fig.~\ref{fig:gammaCombination_D3h} using the 
method of Ref.~\cite{PLUGIN} (see also Refs.~\cite{LHCb-PAPER-2013-020,*LHCb-CONF-2014-004}.)  
The value of $\gamma$ is found to be $(74^{+20}_{-23})^{\rm o}$ for the $\Bm\to DX_s^-$-only fit, and $(74^{+20}_{-19})^{\rm o}$ for 
the for the combined $\Bm\to DX_s^-$ and $\Bm\to DX_d^-$ fit. The value of $r_B$ is nearly identical in the two cases,
with corresponding values of $r_B = 0.081^{+0.025}_{-0.027}$ and $r_B = 0.081^{+0.026}_{-0.029}$.
As expected, most of the sensitivity comes from the $\Bm\to DX_s^-$ decay mode. This value is almost identical
to the LHCb combined result of $(73^{+9}_{-10})^{\rm o}$ found in Ref.~\cite{LHCb-PAPER-2013-020,*LHCb-CONF-2014-004}.  
The value of $r_B$ is similar to the values found in other
$\Bm\to D\Km$ decays~\cite{LHCb-PAPER-2012-001,LHCb-PAPER-2012-055,LHCb-PAPER-2014-017,LHCb-PAPER-2014-041,LHCb-PAPER-2015-014},
but smaller than the value of $0.240^{+0.055}_{-0.048}$~\cite{LHCb-PAPER-2014-028} found in neutral $B$-meson decays.
The strong phase $\delta_B$, averaged over the phase space, peaks at 172$^{\rm o}$ for both fits, but
at 95\% CL all angles are allowed. The constraints on the coherence factor
are relatively weak; while the most likely value is close to 1, any value in the interval $[0,1]$ is allowed
at one standard deviation.

In summary, a $pp$ collision data sample, corresponding to an integrated luminosity of 3.0~fb$^{-1}$, has been used to study the
$\Bm\to DX_s^-$ and  $\Bm\to DX_d^-$ decay modes, where the $D$ meson decays to either the quasi-flavor-specific $K\pi$ final state 
or the $\Kp\Km$ and $\pip\pim$ \CP eigenstates. We observe for the first time highly significant signals in the \CP modes for both the favored
and suppressed $\Bm$ decays, and we also report the first evidence for the ADS DCS $\Bm\to[\Kp\pim]_D\Km\pip\pim$ decay.
We measure the corresponding ADS and GLW observables for the first time in these modes. A fit for
$\gamma$ using only these modes is performed, from which we find $\gamma=(74^{+20}_{-23})^{\rm o}$ for the fit with
only $\Bm\to DX_s^-$, and 
$\gamma=(74^{+20}_{-19})^{\rm o}$ for the combined $\Bm\to DX_s^-$ and $\Bm\to DX_d^-$ fit. Values of $\gamma$ below about $25^{\rm o}$
and larger than approximately $165^{\rm o}$ are not excluded by these modes alone, but are excluded when other modes are 
considered~\cite{LHCb-PAPER-2013-020,*LHCb-CONF-2014-004}.
The precision on $\gamma$ in this analysis is comparable to, or better than, most previous measurements.


\section*{Acknowledgements}

\noindent We express our gratitude to our colleagues in the CERN
accelerator departments for the excellent performance of the LHC. We
thank the technical and administrative staff at the LHCb
institutes. We acknowledge support from CERN and from the national
agencies: CAPES, CNPq, FAPERJ and FINEP (Brazil); NSFC (China);
CNRS/IN2P3 (France); BMBF, DFG, HGF and MPG (Germany); INFN (Italy); 
FOM and NWO (The Netherlands); MNiSW and NCN (Poland); MEN/IFA (Romania); 
MinES and FANO (Russia); MinECo (Spain); SNSF and SER (Switzerland); 
NASU (Ukraine); STFC (United Kingdom); NSF (USA).
The Tier1 computing centres are supported by IN2P3 (France), KIT and BMBF 
(Germany), INFN (Italy), NWO and SURF (The Netherlands), PIC (Spain), GridPP 
(United Kingdom).
We are indebted to the communities behind the multiple open 
source software packages on which we depend. We are also thankful for the 
computing resources and the access to software R\&D tools provided by Yandex LLC (Russia).
Individual groups or members have received support from 
EPLANET, Marie Sk\l{}odowska-Curie Actions and ERC (European Union), 
Conseil g\'{e}n\'{e}ral de Haute-Savoie, Labex ENIGMASS and OCEVU, 
R\'{e}gion Auvergne (France), RFBR (Russia), XuntaGal and GENCAT (Spain), Royal Society and Royal
Commission for the Exhibition of 1851 (United Kingdom).



\addcontentsline{toc}{section}{References}
\setboolean{inbibliography}{true}

\ifx\mcitethebibliography\mciteundefinedmacro
\PackageError{LHCb.bst}{mciteplus.sty has not been loaded}
{This bibstyle requires the use of the mciteplus package.}\fi
\providecommand{\href}[2]{#2}

\newpage


 
\newpage

\centerline{\large\bf LHCb collaboration}
\begin{flushleft}
\small
R.~Aaij$^{38}$, 
B.~Adeva$^{37}$, 
M.~Adinolfi$^{46}$, 
A.~Affolder$^{52}$, 
Z.~Ajaltouni$^{5}$, 
S.~Akar$^{6}$, 
J.~Albrecht$^{9}$, 
F.~Alessio$^{38}$, 
M.~Alexander$^{51}$, 
S.~Ali$^{41}$
% ...

{\footnotesize \it
% ...
$ ^{5}$Clermont Universit\'{e}, Universit\'{e} Blaise Pascal, CNRS/IN2P3, LPC, Clermont-Ferrand, France\\
$ ^{6}$CPPM, Aix-Marseille Universit\'{e}, CNRS/IN2P3, Marseille, France\\
% ...
$ ^{9}$Fakult\"{a}t Physik, Technische Universit\"{a}t Dortmund, Dortmund, Germany\\
% ...
$ ^{37}$Universidad de Santiago de Compostela, Santiago de Compostela, Spain\\
$ ^{38}$European Organization for Nuclear Research (CERN), Geneva, Switzerland\\
% ...
$ ^{41}$Nikhef National Institute for Subatomic Physics, Amsterdam, The Netherlands\\
% ...
$ ^{46}$H.H. Wills Physics Laboratory, University of Bristol, Bristol, United Kingdom\\
% ...
$ ^{51}$School of Physics and Astronomy, University of Glasgow, Glasgow, United Kingdom\\
$ ^{52}$Oliver Lodge Laboratory, University of Liverpool, Liverpool, United Kingdom\\
% ...
}
\end{flushleft}

\end{document}